%% file: MNPT_TCI22_2ndREV.tex
\documentclass[a4paper,journal]{article}
\usepackage{amsfonts}

\usepackage{xcolor}
\usepackage{amsmath}
\usepackage{amssymb}
\usepackage{mathrsfs}
\usepackage{graphicx}
\usepackage{hyperref}

\DeclareMathAlphabet{\dutchcal}{U}{dutchcal}{m}{n}
\SetMathAlphabet{\dutchcal}{bold}{U}{dutchcal}{b}{n}
\DeclareMathAlphabet{\dutchbcal} {U}{dutchcal}{b}{n}

\usepackage{float}
\usepackage{epstopdf}

\usepackage{amsmath,amstext,amssymb,graphicx,hyperref}
\usepackage{verbatim}
\usepackage{color}
\usepackage{algorithm,algorithmic}

\usepackage{mathtools}


\def\dsp{\displaystyle}

\def\vect#1{\mbox{\boldmath{$#1$}}}

\def\bfchi{\mbox{\boldmath$\chi$}}
\def\bfd{\mbox{\boldmath$d$}}
\def\Cc{{\cal C}}
\def\Nc{{\cal N}} 
\def\mC{\mathbb{C}}

\newcommand{\bz}{\vect z}

\newcommand{\cC}{\mathcal{C}}

\def\mR{\mathbb{R}}
\def\mC{\mathbb{C}}

\newcommand{\eps}{\varepsilon}
\renewcommand{\epsilon}{\eps}

\renewcommand{\geq}{\geqslant}

\begin{document}
\title{Quantitative phase and absorption contrast  imaging}

\author{Miguel Moscoso%
\thanks{Department of Mathematics, Universidad Carlos III de Madrid, Leganes, Madrid 28911, Spain.}%
 \and 
 Alexei Novikov%
 \thanks{Mathematics Department, Penn State University, University Park, PA 16802}%
 \and 
George Papanicolaou%
\thanks{Department of Mathematics, Stanford University, Stanford, CA 94305}%
\and 
Chrysoula Tsogka
\thanks{Department of Applied Mathematics, University of California, Merced, 5200 North Lake Road, Merced, CA 95343}}

\maketitle
\begin{abstract}

We present an algorithm for coherent diffractive imaging with phaseless measurements. It treats the forward model as a combination of coherent and incoherent waves. The algorithm reconstructs absorption and phase contrast that quantifies the attenuation and the refraction of the waves propagating through an object. It requires coherent or partially coherent illuminations, and several detectors to record the intensity of the distorted wave that passes through the object under inspection. The diversity of illuminations, obtained by putting masks between the source and the object, provides enough information for imaging. The computational cost of our algorithm is linear in the number of pixels of the image. Therefore, it is efficient for high-resolution imaging. Our algorithm guarantees exact recovery if the image is sparse for a given basis. Numerical experiments in the setting of phaseless diffraction imaging of sparse objects validate the efficiency and the precision of the suggested algorithm.

\end{abstract}


\section{Introduction}
The success of imaging an object with electromagnetic waves is determined as much by the development of new hardware elements 
as by the progress in designing efficient algorithms.  These algorithms solve an inverse problem that, most often, is linear. This is the case, for example, of coherent imaging
when the complex amplitudes of the waves are recorded, or the case of incoherent imaging when only the amplitudes squared, or intensities, are recorded. If, however, waves propagate coherently but only the intensities can be measured, then the imaging problem is nonlinear. This is the so called phase retrieval problem where we seek to reconstruct information about an object from intensity measurements of the wave traversing it.

In this context, we consider multiple measurements of a sparse complex-valued object $\vect f\in\mC^K$, which in our case is the complex refractive index. The  measurements $|\vect b_i|^2\in\mR^\Nc$ of this {\em phase object} of size $K$ are of the form
\begin{equation}
\label{eq:measurements_intro}
\displaystyle
|\vect b_i|^2 = \left| {\cal T}(\vect w_i \circ \vect f) \right|^2\, , \,\,i=1,2,\dots,N,
\end{equation}
where $\vect w_i$  is a  mask, and $\circ$ denotes element-wise multiplication.  In (\ref{eq:measurements_intro}), $ {\cal T}$ is a linear transformation representing the wave propagation operator, which is often approximated by the Fourier transform. 
Although the available measurements  to recover $\vect f$ are intensities, 
the missing phases are still encoded in the  recorded data because the waves propagate coherently and, hence, there is a fixed phase relationship between waves emerging from different points of $\vect f$. Uniqueness of phase retrieval solution requires more than one diffraction pattern, so $N>1$ in \eqref{eq:measurements_intro}.   The different 
diffraction patterns can be obtained using different illumination settings or, equivalently, different masks $\vect w_i$  {in \eqref{eq:measurements_intro}, see} \cite{Balan09,Fannjiang12, Novikov15, Candes15}. In practice, masks are implemented using a spatial light modulator (SLM) or a digital micromirror device (DMD) \cite{Sampsell94,Liu08,Falldorf10, Zheng17}.

We present a two-step algorithm for phase retrieval of sparse objects that mimics the forms in which waves propagate. In the first step, we  assume that the intensities add incoherently, and we treat the coherent contribution to the  data as a modeling error that is absorbed by a denoising algorithm. We use 
a {\em Noise Collector}  \cite{Moscoso20b} as our denoising algorithm.

In the case in which the object is interrogated incoherently, as in for example ghost imaging \cite{Gibson20}, the modeling error is zero, and this step produces exact support recovery if the data are not too noisy. Moreover, if the noise in the data is very strong, the support 
is always contained within the true support.
For this very reason,  only the strong absorbers are recovered in the first step if the object is interrogated coherently. Hence, a second step that takes into account the coherent contribution to the data in (\ref{eq:measurements_intro}) is implemented. With this second step,  the weak (semi-transparent) absorbers are recovered as well. 

Once the support of both the strong and the weakly absorbing structures is known, a straightforward third step that solves the complete problem
is implemented for a precise quantitative phase image restoration. This third step uses both the coherent and the incoherent contributions to the data.

This algorithm has applications in, for example,  phase-contrast imaging that seeks to visualize semi-transparent structures which are otherwise invisible in  conventional absorption-contrast images. The idea behind this imaging modality is simple. When a wave is transmitted through an object it is not only absorbed, but also bent inducing phase changes. The phase changes themselves are invisible, so one needs a method to make them visible as brightness variations in the images. Our algorithm works in the far-field regime, where phase-contrast appears as the result of the free propagation of the waves that transforms phase variations due to the presence of an object into detectable intensity variations in the images \cite{Cloetens96,Nugent96,Wilkins96}. This regime appears to be one of the most  relevant and easy to implement for clinical and biomedical research purposes \cite{Lewis05, Gureyev09}. We refer the reader to \cite{Bravin13} for an extensive discussion on 
phase-contrast imaging in clinical and biomedical applications, and to \cite{Park18} for recent advances in optical phase imaging for investigating cells and tissues in biomedicine. Some methods for phase-contrast imaging that work well in the near-field regime but need specialized optical elements, that are hard to manufacture, are Zernike phase-contrast microscopy \cite{Zernike55}, analyzer-based methods \cite{Davis95a,Davis95b}, and grating-based methods \cite{Pfeiffer06,Pfeiffer07}.

The proposed algorithm also finds applications in fields such as exploration seismology, where phase errors in the data are often strong, so classical algorithms fail  because they try to match the incorrect phase information.  Hence, the measured phases are discarded and the inverse problem is formed for intensity-only data \cite{Gholami14, Aghamiry20}.

The paper is organized as follows. In Section \ref{sec:contributions}, we review some of the most popular algorithms for phase retrieval, and we mention our main contributions. In Section \ref{sec:model}, we introduce our model that produces the measurements for absorption and phase contrast imaging. In Section \ref{sec:nc}, we summarize the construction of the {\em Noise Collector} and some of its important properties. In Section \ref{sec:algo}, we present the algorithm that produces these images. 
Section \ref{sec:numerics} shows the numerical experiments. Section \ref{sec:conclusions}  summarizes our conclusions. 

\section{Related work $\&$ main contributions}
\label{sec:contributions}

There exists an important variety of  algorithms that serve as approximate inverses of the problem stated in \eqref{eq:measurements_intro}. Gerchberg and Saxton \cite{GS72}, and later Fienup \cite{Fienup82}, introduced a scheme based on iterative projections which is simple to implement and proved to be very flexible in practice. These methods are fast but, due to the absence of convexity of problem \eqref{eq:measurements_intro}, they do not always converge to the true solution unless good prior information about the sought object or signal is available. We refer the readers to  \cite{Marchesini07} for more details on these alternating projection techniques.

To guarantee convergence, Chai {\em et al.} \cite{CMP11} and Cand\'es {\em et al.} \cite{Candes13} proposed a different approach that lifts  problem (\ref{eq:measurements_intro}) to a higher dimension to make it convex. The corresponding algorithm solves a low-rank matrix linear system using nuclear norm minimization. The algorithm converges to the true solution, even without prior information about the object and independently of its complexity, provided the recorded data is diverse enough. 

However, when lifting the problem to a higher dimension,  the size increases quadratically in the number of unknowns {$K$}, and the solution becomes infeasible for $K$  large.

The authors in \cite{Waldspurger15}  follow a similar approach. They reformulate  phase recovery as 
a {\em MaxCut}-like semidefinite program, and solve the resulting problem using a block coordinate descent algorithm similar to the    Gerchberg and Saxton algorithm  \cite{GS72}. Although with this approach the problem can be solved more efficiently, 
its size also increases quadratically, so the approach is not suitable for large scale 
problems. 
Other interesting algorithms that require matrix lifting are, for example, \cite{Shechtman11,Ohlsson12}.

Inspired by advances in compressed sensing, and in order to improve the efficiency of the existing algorithms, several works explore the idea of sparsity as prior information on the sought objects or signals, with sparsity levels $\dutchcal{s}\ll K$. In \cite{Shechtman14}, for example, the authors propose an efficient local search method that empirically recovers  $\dutchcal{s}$-sparse objects, with $\dutchcal{s}\sim O(\Nc^{1/3})$. Here $\Nc$ stands for the dimension of the data.  However, convergence to the correct solution is not guaranteed. In \cite{Jaganathan17},   the authors exploited the fact that  the one-dimensional Fourier phase retrieval problem is equivalent to the {\it turnpike reconstruction problem}, where the objective is to reconstruct a set of vertices from the set of their pairwise distances. They showed that $O(\dutchcal{s}^4)$
reconstruction times are possible when the set of  vertices is $\dutchcal{s}$-sparse and $\dutchcal{s}\sim o(\Nc^{1/2})$.  In  \cite{Novikov21}, the authors generalize  the one-dimensional algorithm~\cite{Jaganathan17} to higher dimensions and obtained  dramatically
faster $O(\dutchcal{s}^2)$  reconstruction times of  $\dutchcal{s}$-sparse objects, with $\dutchcal{s}\sim o(\Nc^{1/2})$. These algorithms do solve efficiently high-dimensional sparse problems.  However, they only work for Fourier phase retrieval settings, i.e. ${\cal T}$ must be a Fourier transform in \eqref{eq:measurements_intro}.

In this paper, we present an efficient and robust phase retrieval algorithm for $O(\Nc^{1/2})$-sparse objects, whose computational cost grows linearly with the problem size $K$, so it is suitable for large scale problems. Although for ease of presentation it is introduced for Fourier measurements, it also works for general quadratic measurements without any modification. The algorithm
vectorizes the matrix formulation in \cite{CMP11} and \cite{Candes13} to solve an optimization vector problem, and uses a {\em Noise Collector} \cite{Moscoso20b} to reduce its dimensionality \cite{Moscoso21}.

The huge reduction of dimensionality is carried out in two steps mimicking the forms in which waves travel. In this way, the algorithm merges coherent and incoherent imaging, as it considers the coherent and incoherent contributions to the data sequentially. With this approach we are able to image transparent or semi-transparent structures that are not visible in the commonly used absorption-based images.

As a by-product, our algorithm can also be used for phase retrieval with partially coherent observations without any modification and without loss of resolution. This might be important for applications where fully coherent sources are hard to produce as in,  for example, X-ray phase-contrast imaging \cite{Jingshan14}. In these cases, only the signal-to-noise ratio (SNR) of the created images is affected. In the extreme case in which the observations are fully incoherent, the transparent structures cannot, of course, be visualized.

\section{Model}
\label{sec:model}
 
When a wave propagates through an object, both its amplitude and phase are altered. 
Intuitively, the amplitude of the transmitted wave depends on the absorption of the wave, 
while its phase shift depends on the refraction. 

Consider a planar object of finite support and thickness $l$  illuminated perpendicularly by a monochromatic plane wave of wavenumber $k=2\pi/\lambda$ traveling in the $z$ direction, { see Figure~\ref{fig0}}. If the wavelength $\lambda$ is small compared to its
dimension $l$, the interaction of the wave with the object can be described in terms of integrals of the
complex refractive index $n(x,y)$, so its  transmissivity is given by
\begin{equation}
\label{eq:transmissivity0}
t(x',y') =  \displaystyle e^{{i k \int_{l} n(x',y') dz'}}\, .
\end{equation} 
Then, the diffracted complex amplitude in the far-field, using the Fraunhofer
approximation, is 
\begin{equation}
\label{eq:fourier2d}
b(x,y) = - i \frac{e^{ikL}}{\lambda L}\int t(x',y') e^{i2\pi(xx'+yy')}dx'dy'\, ,
\end{equation}
where $L\gg 1$ is the distance between the object plane and the detector plane. We use prime symbols for the coordinates in the object plane to avoid confusion with the $(x,y)$ coordinates in the detector plane. We use dimensionless coordinates by scaling the transverse coordinates $(x,y)$ with $\sqrt{\lambda L}$.
Fraunhofer diffraction occurs when $L\gg a^2/\lambda$, where $a$ is a characteristic transverse length of the object. Physically, this means that the object is far enough from the sources, so the incident waves are effectively plane waves, i.e., the phase of the waves at each point on the object is the same.

Mathematically, the complex refractive index in (\ref{eq:transmissivity0}), is expressed as
\begin{equation}
\label{eq:rindex}
n(x',y')=1 - \delta(x',y')+ i\beta(x'y')\,.
\end{equation}
It is the ratio of the wavenumber in the object $\tilde k(x',y')$ and the wave number in the vacuum $k$ and, hence, is a measure of how fast the waves travel through the object. 
The real and imaginary components
${\displaystyle  \delta }$ and ${\displaystyle \beta}$ determine the refraction and absorption effects of the interaction wave-matter, respectively.
For a thin planar object, the phase shift and the absorption coefficient are well approximated by 
\begin{eqnarray}
\label{eq:phaseshift}
\phi(x',y') &=& k \int_{l}\delta(x',y') dz' \,\quad \mbox{and}\\
\label{eq:absorption}
\mu(x',y') &=& 2 k \int_{l}\beta(x',y') dz' \,, 
\end{eqnarray}
respectively. Here, the integration is taken over the extend of the object ${l}$ along the direction of wave propagation $z$. Both \eqref{eq:phaseshift} and  \eqref{eq:absorption} are proportional
to the density of electrons at each point $(x',y')$ of the object and, hence, both allow for reconstructions of electron densities. 
However, these reconstructions are usually based on wave attenuation only, as phase shifts still remain harder to quantify. 

However, phase shifts also induce intensity variations on the images, and this effect can reveal important features of the object's structure that are not visible in the attenuation-based images. This is of great importance at high energy ranges where wave attenuation is  small, or even negligible. Indeed, the refractive index decrement  $\delta$ and the extinction coefficient $\beta$ have a strong dependence upon  the energy of the incident wave $E$, but they behave very differently as the energy increases. In particular,
for X-rays, $\delta$ decreases approximately as $1/E^2$, while $\beta$ approximately as $1/E^4$, so at these energies $\delta$ is between one and three orders of magnitude larger than $\beta$.  For example, at $E=30 keV$, $\delta\approx 2.56 10^{-7}$ and $\beta\approx 1.36 10^{-10}$ for water.

\subsection{Weak phase-contrast}

If the absorption is negligible so $\beta(x',y')\ll \delta(x',y')$, then the phase shift (\ref{eq:phaseshift})  modulates the detected complex amplitude (\ref{eq:fourier2d}), with transmissivity  given by 
\begin{equation}
\label{eq:transmissivity}
t(x',y') = e^{i\phi(x',y')}\, .
\end{equation}
If, in addition,
$\phi(x',y')$ is small so $\beta(x',y')\ll \delta(x',y')\ll 1/l$, (\ref{eq:transmissivity}) can be approximated as
\begin{equation}
\label{eq:transmissivity_approx}
t(x',y') \approx 1 + i \phi(x',y')\,,
\end{equation}
and the phase-contrast is weak. In this case, the diffracted complex amplitude is given by
\begin{equation}
\label{eq:weak}
b_{tot}(x,y) = - i \frac{e^{ikL}}{\lambda L} \int [1 + i \phi(x',y')] e^{i2\pi(xx'+yy')}dx'dy'\, .
\end{equation}
This is called the weak phase object approximation.
Integration of the first term gives a delta function (for an infinite aperture) that represents the direct wave that goes through the object without interaction. We do not consider it here, as it is usually blocked by a beam stop \cite{Veen04}. Thus, only the scattered component 
\begin{equation}
\label{eq:phasecontrast}
b(x,y) = \int \phi(x',y') e^{i2\pi(xx'+yy')}dx'dy'\,,
\end{equation} 
up to a constant that is set here equal to one, is considered for imaging.

Assume that, for imaging purposes, the inspected object is discretized using a grid of $K$ pixels, so its phase $\phi(x',y')$ is well approximated by the size $K$ vector 
\begin{equation}\label{eq:rho}
\vect \phi=[\phi_{1},\ldots,\phi_{K}]^{\intercal}\in\mR^K\, .
\end{equation}
According to (\ref{eq:phasecontrast}), if the object is illuminated by a monochromatic coherent plane wave, 
the complex diffraction pattern $\vect b$ measured at an {${\cal N}$}-pixel detector is given by the  size {${\cal N}$} discrete Fourier transform of the object's phase given by
$$
\vect b = F \vect \phi\, ,
$$
where 
${\displaystyle F_{sk} ={\frac {e^{i2\pi(s-1)(k-1)/K}}{\sqrt {K}}}}$,  with {$s=1,\dots,{\cal N}$}, and $k=1,\ldots ,K$. An important special case is when ${{\cal N}}=K$, so the transformation is the classical discrete  Fourier transform. If ${{\cal N}}<K$, the case considered here, the Fourier transform is said to be {\em undersampled}.

If a known set of $N$ spatially structured patterns or masks
$$
\vect w_i=[w_{i1},\ldots,w_{iK}]^{\intercal}\in\mC^K\,, \quad i=1,\dots,N,
$$ 
illuminate the object, then several complex diffraction patterns 
$$
\vect b_i = F W_i \vect \phi\, , \quad i=1,\dots,N,
$$
are available for imaging. Here, $W_i$ is the diagonal matrix $W_i=\mbox{diag}(w_{i1},\ldots,w_{iK})$ corresponding to the $i$-th illumination pattern. 
With this notation, the $s$-th component  of the data vectors 
\begin{eqnarray}
\label{response0}
(\vect b_i)_s = \sum_{k=1}^K  F_{sk}w_{ik} \phi_k \, , \quad {s=1,\dots, {\cal N},}
\end{eqnarray}
represents the complex field at the $s$-th detector when $i$-th illumination impinges on the object.
If the detectors can only measure intensities, then the problem is to recover (\ref{eq:rho}) from measurements of the type
\begin{eqnarray}
\label{response:intensity0}
|(\vect b_i)_s|^2 &= &\left|\sum_{k=1}^K    F_{sk}w_{ik} \phi_k \right|^2 \nonumber \\
\hspace*{-1.5cm} &\hspace*{-1.5cm}& \hspace*{-2cm} = \sum_{k=1}^K  |w_{ik}|^2 |\phi_k|^2 +
\sum_{k=1}^K \sum_{\substack{k'=1\\k'\neq k}}^K  F_{sk} F_{sk'}^*w_{ik} w_{ik'}^* \phi_k\phi_{k'}\, ,
\end{eqnarray}
$i=1,\dots,N,\, {s=1,\dots, {\cal N}}$. Note that, in this case, the unknown vector $\vect \phi$ is real. The left side represents the  intensity received 
at the $s$-th detector when the
object is illuminated with the $i$-th illumination pattern $\vect w_i=[w_{i1},\ldots,w_{iK}]^{\intercal}$. The first term in the right side of this expression is the incoherent contribution to this intensity, and the second term is the coherent contribution. 
The coherent contribution is characterized by a fixed phase relationship between the waves emerging from different points of the object, 
while in the incoherent contribution this phase relationship does not exist. 
The coherent contribution produces the interferences used in coherent imaging to determine the object's structure, and it is the term
that makes  phase-retrieval  non-linear.

\subsubsection{Incoherent imaging}
According to the previous discussion, if the phases of the waves immediately exiting the object are randomized either by a diffuser or by the medium's inhomogeneities, so  the wavefront is totally scrambled, then the  second term in (\ref{response:intensity0}) is negligible and the inverse problem is linear
in the phase-shifts (squared) $ |\phi_k|^2$. In this case, the intensities received at all the detectors are equal; the data vector does not depend on $s$. Thus, to improve the SNR, the collected intensities  can be averaged and we can form the linear system
\begin{equation}
\label{eq:incohsyst}
  \underbrace{\begin{bmatrix}
   |w_{11}|^2 & |w_{12}|^2 & \hdots & |w_{1K}|^2\\
   |w_{21}|^2 & |w_{22}|^2 & \hdots & |w_{2K}|^2  \\
    \vdots &  \vdots &                   & \vdots \\ 
    |w_{ N1}|^2 & |w_{N 2}|^2 & \hdots & |w_{N K}|^2
  \end{bmatrix}}_{ \displaystyle {\cal W }_{incoh} }
  \underbrace{\begin{bmatrix}
    |\phi_1|^2  \\
    |\phi_2|^2  \\
    \vdots \\
    |\phi_K|^2
  \end{bmatrix}}_{\vect \chi} =
  \underbrace{ \begin{bmatrix}
    d_1  \\
    d_2  \\
    \vdots \\
    d_{N}
  \end{bmatrix}}_{\vect d_{avg}}\, .
\end{equation}
Here, each row of the matrix ${\cal W}_{incoh}$ contains the intensities upon the pixelated object produced by the $N$  masks $\vect w_i$. The linear system
\begin{equation}
{\cal W}_{incoh} \,\, \vect \chi = \vect d_{avg}
\end{equation}
can be easily solved by means of any $\ell_1$- or $\ell_2$-method depending on the number of masks $N$
used for imaging and the sparsity of the vector  $\vect \chi$, which is in this case real.

\subsection{Non-weak phase-contrast}
If the absorption is negligible but the phase contrast is not weak but strong, then (\ref{eq:transmissivity}) cannot be approximated by 
(\ref{eq:transmissivity_approx}) and we should image the transmissivity vector
\begin{equation}\label{eq:transmissivity_vector}
\vect t=[t_1,\ldots,t_K]^{\intercal}=[e^{i\phi_{1}},\ldots,e^{i\phi_{K}}]^{\intercal}\in\mC^K\, ,
\end{equation}
which is now complex. Then, (\ref{response:intensity0}) becomes
\begin{eqnarray}
\label{response:intensity1}
|(\vect b_i)_s|^2 &=& \left|\sum_{k=1}^K    F_{sk}w_{ik} t_k \right|^2 \nonumber \\
&&\hspace*{-1.3cm} = \sum_{k=1}^K  |w_{ik}|^2 +
\sum_{k=1}^K  \sum_{\substack{k'=1\\k'\neq k}}^K F_{sk} F_{sk'}^*w_{ik} w_{ik'}^* t_kt_{k'}^*\, ,
\end{eqnarray}
$i=1,\dots,N,\, {s=1,\dots, {\cal N}}$. In \eqref{response:intensity1}, the first term in the right side is the total intensity $\vect c$   upon the object, which is known, and that does not provide any information for imaging. All the information 
is contained in the interferences of  the scattered wave represented by the second term. 

At least two approaches are possible to solve the nonlinear problem (\ref{response:intensity1}). One is to solve it iteratively for the $K$ complex unknowns $t_i$ as in \cite{GS72, Fienup82}. This phase retrieval algorithms work well in practice if good prior information about the object of interest is known; otherwise, convergence to the true solution is not guaranteed. The second option is to reformulate (\ref{response:intensity1})  as a linear  matrix problem  for the $K^2$ unknowns $t_it_j^*$, $i,j=1,\dots,K$, as in \cite{CMP11, Candes13}, and solve the resulting problem by using nuclear norm minimization. This option guarantees convergence to the unique solution without any prior information about the sought object, but the number of unknowns grows quadratically with the number of unknowns $K$ and, thus, the solution becomes unfeasible for high-resolution images with large $K$. 

Alternatively, one can define the cross-correlation
vector of $\vect t$ 
\begin{equation}\label{eq:crossrho}
\vect \chi_{cross}=[t_{1}t_{2}^*,t_{1}t_{3}^*,\ldots,t_{1}t_{K}^*,t_{2}t_{1}^*,t_{2}t_{3}^*,\ldots,t_{2}t_{K}^*,t_{3}t_{1}^*,\ldots,]\, ,
\end{equation}
excluding the real terms $|t_{i}|^2=1$,  
and solve the huge linear system 
\begin{equation}
\label{eq:hugelinearsystem}
 {\cal W}_{coh}\,\,\vect \chi_{cross} = \vect d\, ,
\end{equation}
given the data 
$\vect d = [\vect d_1^{T}, \dots, \vect d_{\cal N}^{T}]^T$. 
Here, $\vect d_s = [|(\vect b_1)_s|^2,\dots,|(\vect b_N)_s|^2]^T  - \vect c $  represents the $N$ intensities recorded at the detector $s$ minus  the total intensity  upon the object $\vect c$. 
In \eqref{eq:hugelinearsystem},
\begin{equation}
\label{eq:coh_matrix}
 {\cal W}_{coh} = [({\cal W}_{1,coh})^{T}, ({\cal W}_{2,coh})^{T}, \dots, ({\cal W}_{{\cal N},coh})^{T}]^T
\end{equation}
is a huge matrix of size {${\cal N} N  \times K(K-1)$}, where ${\cal W}_{s,coh}$ is defined in (\ref{eq:coh_matrix_n})
\begin{table*}[htbp]
\begin{equation}
\label{eq:coh_matrix_n}
{\cal W}_{s,coh}=
  \begin{bmatrix}
  c_{s1,12} & c_{s1,13} & \hdots & c_{s1,1K} & c_{s1,21} & c_{s1,23} & \ldots & c_{s1,2K} & c_{s1,31} &\ldots \\
 c_{s2,12} & c_{s2,13} & \hdots & c_{s2,1K} & c_{s2,21} & c_{s2,23} & \ldots & c_{s2,2K} & c_{s2,31} &\ldots \\
    \vdots &  \vdots &      \ldots &  \vdots  &\vdots &  \vdots              & \ldots &  \vdots              & \vdots &  \ldots              \\ 
   c_{sN,12} & c_{sN,13} & \hdots & c_{sN,1K} & c_{sN,21} & c_{sN,23} & \ldots & c_{sN,2K} & c_{sN,31} &\ldots \\
  \end{bmatrix}\, ,
\end{equation}
\end{table*}
with $c_{si,lm}=F_{sl} F_{sm}^*w_{il}w_{im}^*$. This matrix models the coherent component of the intensities received
at the $s$-th detector corresponding to the $N$ illumination patterns $\vect w_i$ used to {\em filter} the object. If the resolution is low, so the number of pixels $K$ is small, one could easily solve (\ref{eq:hugelinearsystem}) using an appropriate solver. However, as in the approach suggested in \cite{CMP11, Candes13}, the size of the problem  increases quadratically with $K$, so the search of a solution rapidly becomes prohibitive for large values of $K$, as well. 

\subsection{Absorption and phase contrast}
Absorption can be easily taken into account by using  (\ref{eq:phaseshift}) and (\ref{eq:absorption}) in (\ref{eq:transmissivity0}). If absorption is not negligible, then the transmissivity is given by
\begin{equation}
\label{eq:transmissivity1}
t(x',y') = e^{-\mu(x',y')/2} e^{i\phi(x',y')} = |t(x',y')| e^{i\phi(x',y')}\, .
\end{equation}
This is the most general case in which absorption and refraction effects are mixed in the images. In these cases, the transmissivity vector is given by the complex vector
\begin{equation}\label{eq:transmissivity_vector_absorption}
\vect t=[t_1,\ldots,t_K]^{\intercal}=[|t_1|e^{i\phi_{1}},\ldots,|t_k|e^{i\phi_{K}}]^{\intercal}\in\mC^K\, ,
\end{equation}
with amplitudes different than $1$. Then, the problem is to find (\ref{eq:transmissivity_vector_absorption}) from measurements of the form
\begin{eqnarray}
\label{response:intensity2}
|(\vect b_i)_s|^2 &=& \left|\sum_{k=1}^K    F_{sk}w_{ik} t_k \right|^2 \nonumber \\
&&\hspace*{-1.8cm}= \sum_{k=1}^K  |w_{ik}|^2 |t_k|^2 +
\sum_{k=1}^K \sum_{\substack{k'=1\\k'\neq k}}^K  F_{sk} F_{sk'}^*w_{ik} w_{ik'}^* t_k t_{k'}^*\,  ,
\end{eqnarray}
{$i=1,\dots,N,\, s=1,\dots, {\cal N}$}. This problem has, of course, the same form as before, but now the incoherent contribution to the intensity is modulated by the absorption. This makes attenuation-based imaging possible with incoherent sources.

The main purpose of this paper is to propose a dimension reduction technique 
that allows  to find the solution of  (\ref{response:intensity2}) efficiently for large $K$.  This algorithm is presented in Section \ref{sec:algo}. 
The set of equations  \eqref{response:intensity2} can be written in matrix form as
\begin{equation}
\label{eq:hugelinearsystem_absorption}
  {\cal W}_{incoh}\,\,\vect \chi + {\cal W}_{coh}\,\,\vect \chi_{cross} = \vect d\, ,
\end{equation}
with ${\cal W}_{incoh}$ and ${\cal W}_{coh}$ defined in \eqref{eq:incohsyst} and \eqref{eq:coh_matrix}, respectively. Solution of (\ref{eq:hugelinearsystem_absorption}) provides {\em complete} images, with  distributions of both
the real and imaginary parts of the complex refractive index \eqref{eq:rindex}. 
However, the bottleneck is still the size of the problem, which is enormous if one wants to form high resolution images. An image with only $1000\times 1000$ pixels, amounts to solving a linear system with $10^{12}$ unknowns!

In Section \ref{sec:algo}, we describe the proposed algorithm that makes possible to find the desired solution in polynomial-time if the  object to be image is sparse in some appropriate basis, but first we summarize the main properties and the construction of the {\em Noise Collector} which is an essential denoising tool in this approach. For simplicity, we will assume that the sought object is sparse in the real space, meaning that only a few pixels in the object plane absorb or bend the waves. We note, though, that many images are naturally compressible by using appropriate sparsifying transforms such as wavelets, or other dictionaries that are directly adapted to the data \cite{Ravishankar13}.

\section{The noise collector}
\label{sec:nc}
The {\em Noise Collector}~\cite{Moscoso20b} is a denoising algorithm to find  the vector $ {\mbox{\boldmath$\chi$}}  \in \mathbb{C}^{\cal K}$ in
\begin{equation}
\label{eq:nc_general}
 {\cal A} \, \vect \chi = \vect d_0 + \vect e \,,
\end{equation}
from highly incomplete measurement data $\vect d = \vect d_0 + \vect e \in \mC^{\cal N}$ 
corrupted by additive noise  $\vect e \in \mC^{\cal N}$, where $ 1 \ll {\cal N} < {\cal K}$.  Here, ${\cal A}$ is a general measurement matrix of size
${\cal N}\times {\cal K}$, whose columns have unit length. The main result in~\cite{Moscoso20b} ensures that we can 
recover the support of  $\vect \chi$ by looking at the support of  $\vect \chi_{\tau}$ found as 
\begin{equation}\label{rho_tt}
\begin{array}{c}
\left( \vect \chi_{\tau}, \vect \eta_{\tau} \right) = 
\arg\min_{ \small \vect \chi, \small \vect \eta} \left(    \tau  \| \vect \chi \|_{\ell_1} +  \| \vect \eta \|_{\ell_1}  \right),\\
 \hbox{ subject to } {\cal A} \vect \chi + \Cc \vect \eta =\vect d.
 \end{array}
\end{equation}
Here,  $\tau$ is an  $O(1)$ no-phantom weight, and $\cC \in \mC^{{\cal N}\times \Sigma}$  is a {\it Noise Collector} matrix,  with $\Sigma = {\cal N}^\beta$, for  $\beta>1$ (typically close to one).
If  the noise $\vect e$ is Gaussian, then
the columns of $\Cc$ can be chosen independently and at random on the unit sphere  
 $\mathbb{S}^{{\cal N}-1}$. 
 The weight $\tau>1$ is chosen so it is expensive to approximate $\vect e$ with the columns
of $T$, but it cannot be taken too large because then we lose the signal $\vect \chi$ that gets absorbed by the {\it Noise Collector} as well. 
For practical purposes, $\tau$ is chosen as the minimal value for which $\vect \chi = 0$
when the data is pure noise, i.e., when $\vect d_0=0$. The key property is that the optimal value of $\tau$ does not depend on the level of
noise and, therefore, it is chosen in advance, before the {\it Noise Collector} is used for a specific task.

It can be shown 
that if the matrix ${\cal A}$ is incoherent enough, so its columns are not almost parallel, 
the minimizer in (\ref{rho_tt}) has no false positives for any level of noise, with probability that tends to one as the dimension of the data increases to infinity. 
To find the solution of (\ref{rho_tt}), we  define the function
\begin{eqnarray} 
 F(\bfchi, \vect \eta, \bz) &=& \lambda\,(\tau \| \bfchi \|_{\ell_1} +  \| \vect \eta \|_{\ell_1}) \\
&+& \frac{1}{2} \| {\cal A}  \bfchi  + \Cc  \vect \eta - \bfd \|^2_{\ell_2} + \langle \bz, \bfd - {\cal A} \bfchi - \Cc  \vect \eta \rangle \nonumber
\end{eqnarray}
and determine the solution as
\vspace{-0.2cm}
\begin{equation}\label{min-max}
\max_{\bz} \min_{\bfchi,\vect \eta} F(\bfchi,\vect \eta,\bz) .
\end{equation}
This strategy finds the minimum in (\ref{rho_tt}) exactly for all values of the regularization parameter $\lambda$. Thus,
the  method is fully automated, meaning that it 
has no tuning parameters. To determine the exact extremum in (\ref{min-max}), we use the iterative soft thresholding algorithm GeLMA~\cite{Moscoso12}
 that works as follows.

Pick a value for the no-phantom weight $\tau$; for optimal results calibrate $\tau$ to be the smallest value for which $\vect \chi=0$ when the algorithm is fed with pure noise.  
In our numerical experiments we use $\tau= 2$.
Next, pick a value for the regularization parameter, for example $\lambda=1$, and choose step sizes $\Delta t_1< 2/\|[{\cal A} \, | \, \Cc]\|^2$ and 
$\Delta t_2< \lambda/\|{\cal A} \|$\footnote{Choosing two step sizes instead of the smaller one $\Delta t_1$ improves the convergence speed.}. Set $\vect \bfchi_0= \vect 0$, $\vect \eta_0=\vect 0$, $\vect z_0=\vect 0$, and
iterate for $k\geq 0$:
\begin{eqnarray} 
&& \vect r = \bfd - {\cal A}  \,\bfchi_k - \Cc \,\vect\eta_k\nonumber \, ,\\
&&\vect \bfchi_{k+1}=\mathcal{S}_{ \, \tau \, \lambda \Delta t_1} ( \bfchi_k +\Delta t_1 \, {\cal A} ^*(\vect z_k+ \vect r))
\nonumber \, ,\\
&&\vect \eta_{k+1}=\mathcal{S}_{\lambda \Delta t_1} ( \vect\eta_k +\Delta t_1  \, \Cc^*(\vect z_k+ \vect r))
\nonumber \, ,\\
&&\vect z_{k+1} = \vect z_k + \Delta t_2 \, \vect r \label{eq:algo}\, ,
\end{eqnarray}
where  $\mathcal{S}_{r}(y_i)=\mbox{sign}(y_i)\max\{0,|y_i|-r\}$. 
Terminate the iterations when the distance $\|\vect \bfchi_{k+1} - \vect \bfchi_{k} \|$ between two consecutive iterates
is below a given tolerance.

For more details on the theory and properties of the {\em Noise Collector}, we refer the reader to \cite{Moscoso20b, Moscoso21}.

\section{Dimension reduction}
\label{sec:algo}
Instead of solving the linear system with $K^2$ variables (\ref{eq:hugelinearsystem_absorption}) 
for non-weak phase objects, with or without absorption, we propose to reduce its dimensionality by constructing a linear problem for only $O(K)$ significant unknowns, and absorb the error corresponding to the contribution of the unmodeled unknowns by using a {\em Noise Collector}. Mathematically, we propose to solve the linear system 
\begin{equation}
\label{eq:nc}
{\cal A} \vect \chi + {\cal C}\vect \eta = \vect d\, ,
\end{equation}
where ${\cal A}$ is a matrix with $O(K)$ subsampled columns of the matrix $[ {\cal W}_{incoh} | {\cal W}_{coh}]$, each column of size {${\cal N} N$}. In this formulation, $\vect \chi $ is a sparse vector that represents the object, $\vect \eta$ is an unwanted vector with no physical meaning that absorbs the noise, and $\cal C$ is a {\em Noise Collector} matrix  with ${\cal N}^\beta$ columns drawn independently and at random on the unit sphere
(we use $\beta=1.5$ in our simulations).

The algorithm has three steps. 
\begin{itemize}
\item[(1)] In the first step, we seek the strong absorbing objects. We set ${\cal A}={\cal W}_{incoh}$ so ${\cal W}_{coh}=0$, and solve \eqref{eq:nc}  for $\vect \chi=[|t_1|^2,|t_2|^2,\dots,|t_k|^2]^T$. The term 
${\cal C}\vect \eta$ in \eqref{eq:nc}, where  $\cal C$ is a small matrix with ${\cal N}^\beta$ columns, absorbs the coherent contributions to the intensities that are treated in this step as noise. Since the model we solve is not exact, only the strong absorbing objects are detected.
\item[(2)] In the second step, we seek the non-absorbing objects. Since these objects are almost transparent and do not have a significant impact on the recorded intensities, we look for their phases, encoded in the vector $\vect \chi_{cross}$ defined in \eqref{eq:crossrho}. To this end, we first subtract the incoherent contribution to the recorded intensities due to the detected objects in the first step. Then, 
we set ${\cal W}_{incoh}=0$ and ${\cal A}=({\cal W}_{coh})_{sub}$, where $({\cal W}_{coh})_{sub}$ is a small subsampled matrix of the huge matrix ${\cal W}_{coh}$. It only contains the $m\,(K-1)$ columns that correspond to the interactions between the $m$ detected objects in the first step and the other pixels in the image. Since we are not modeling the incoherent contributions of the remaining $(K-m)$ pixels, the system we solve is not exact neither. Hence, we also use a Noise Collector matrix $\cal C$ with ${\cal N}^\beta$ columns to absorb the noise.
\item[(3)] The third step is optional. It is used to obtain more precise quantitative images. Once the strong and weakly absorbing objects are found, we solve the full problem \eqref{eq:hugelinearsystem_absorption} restricted to the recovered support. This is now a small problem that can be solved using an $\ell_2$ minimization method that gives very accurate results.
\end{itemize}

We stress that, for this dimension reduction strategy, it is necessary that the unknown object can be represented as a sparse vector. Otherwise, the (modeling) errors are too big to be absorbed.

\section{Numerical experiments}
\label{sec:numerics}
\begin{figure}[htbp]
\centerline{\includegraphics[width=0.7\textwidth]{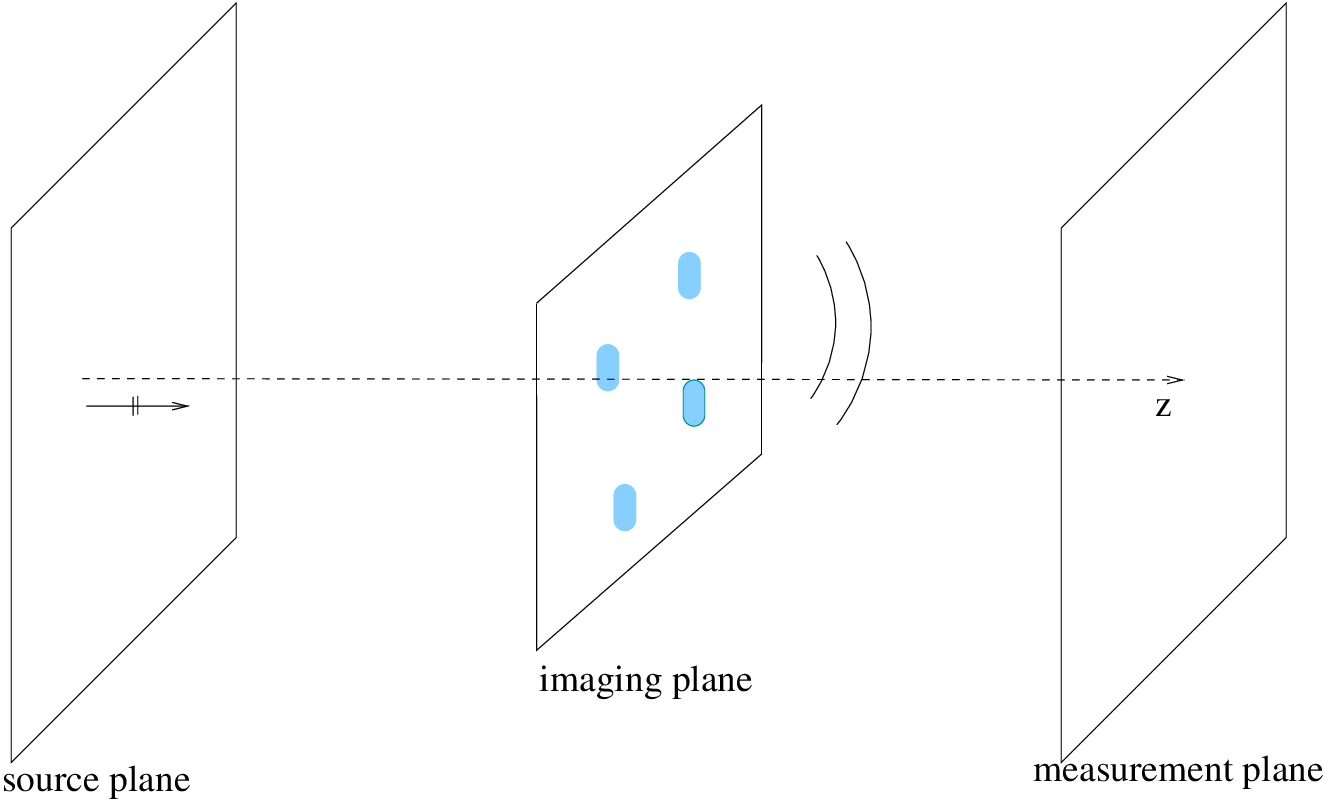}}
\caption{Schematic of the imaging setup. We seek to reconstruct the transmissivity vector $\vect t$ on the imaging plane by recording the medium's response on the measurement plane when $N$ known excitations are sent from the source plane. {The transmissivity vector $\vect t$ is a complex-valued vector of dimension equal to the pixels used to discretize the imaging plane; the areas where the transmissivity is different  from one}  are illustrated in the schematic with light blue color.} 
 \label{fig0} 
 \end{figure}

The simulations shown here illustrate the potential of the proposed algorithm. 
In this work, a thin object is illuminated with quasi-monochromatic coherent or with partially incoherent sources. The schematic for the imaging setup is shown in Figure \ref{fig0}. The units of the problem are given with respect to the central wavelength 
$\lambda_0$, as this is the important parameter in the simulations, and all other length scales will be referred to it.  The main assumption is the sparsity of the object. We show here reconstructions of small point-like structures. As for other compressed sensing based algorithms the methodology can be used for more complex imaging scenes, as long as there is a sparsifying transform that allows a sparse representation of the scene. In some applications, they use an off-the-shelf transform like the Fourier, Hadamard, wavelet, or curvelet, while in others they find a new transform using dictionary learning. 
Note that the sparsity is unknown and can scale as $\sqrt{{\cal N} N}$ with ${\cal N} N$ the number of data (intensities) measured.   

The sources are located on a two dimensional array of size $16000 \lambda_0 \times 16000 \lambda_0$,  
at a distance of $16000 \lambda_0$ 
from the object. There are $21 \times 21$ point sources, evenly distributed on the source plane. These are used to create $N$ illumination patterns on the imaging plane. The illuminations are created by assigning random amplitudes and phases to each one of the point sources. The resulting illumination patterns $\vect w_i$, for $i=1,\ldots,N$, are assumed to be known on the imaging plane. 

We use  $5 \times 5$ receivers $16000 \lambda_0$ downstream to collect the data corresponding to $N=300$ different illumination patterns. The receivers are evenly distributed on the measurement plane that is parallel to the object and to the transmitting array (see Figure \ref{fig0}). The receiving array also has an aperture of $16000 \lambda_0 \times 16000 \lambda_0$. 
Wave propagation is modelled using the 3D wave Green's function, 
$$ G(\vect x, \vect y) = \frac{e^{i k |\vect x -\vect y|}}{ 4 \pi |\vect x -\vect y|} $$
with $|\vect x -\vect y|$ the distance between points $\vect x $ and $\vect y$ and $k_0=2\pi/\lambda_0$ the wavenumber. 
 
In order to form an image, the imaging plane is discretized using $31\times 31$ pixels, with pixel size equal to half a wavelength in both directions., {\em i.e.} a square of size $\lambda_0/2 \times \lambda_0/2$.  Although we consider a typical transmission setup, same results are obtained for reflection or for more complicated sources and receivers layouts. 

We consider absorbing and non-absorbing objects that change the phases of the waves that go through them.  { We form images of  $\rho(x',y')=t(x',y') -1$. 
The illumination by the direct wave may be absorbed by adding a column to the Fourier transform matrix $F$. 
 In our numerical experiments, 
$\rho_k = e^{i\pi}$ if there is a strong absorbing object at pixel $k$, and $\rho_k = 0.1 \, e^{i\pi/2}$ if there is a non-absorbing object.} If only  intensities are recorded, these non-absorbing objects are very hard to image because the waves that go through them change only slightly.

\begin{figure}[htbp]
\begin{center}
 true $| \rho_k |^2$ \\ 
 \includegraphics[scale=0.24]{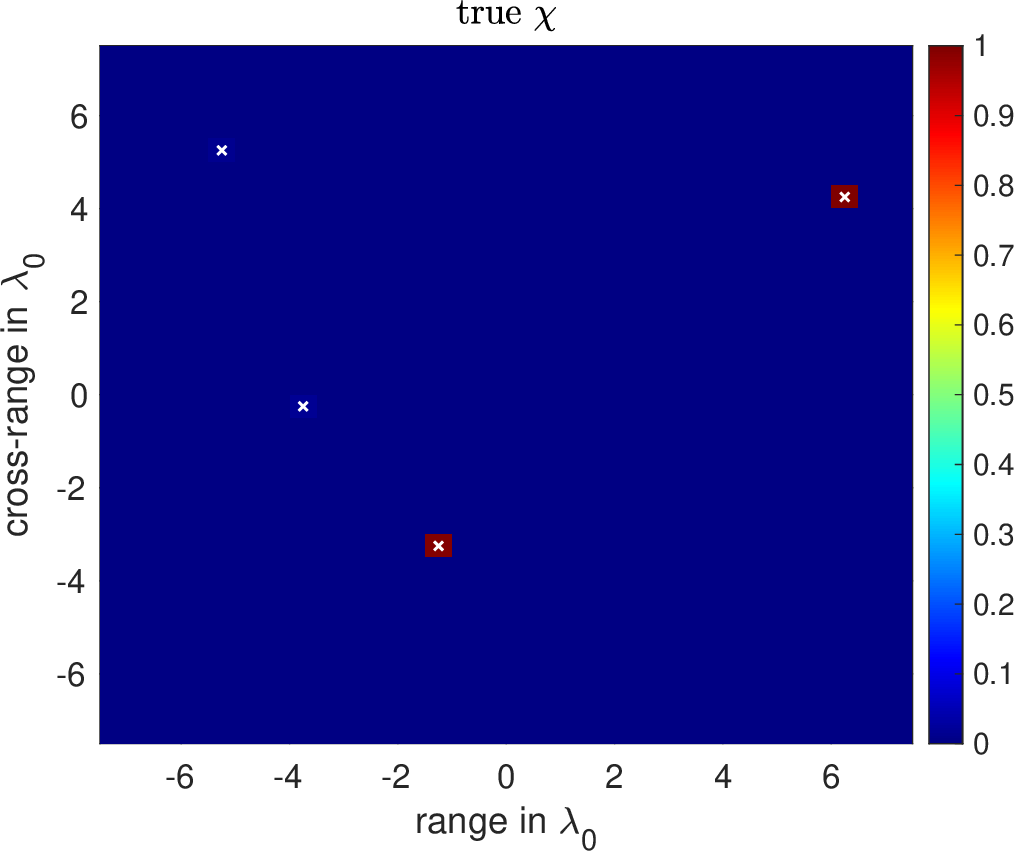} \\
  \begin{tabular}{cc}
 $\ell_1$ no noise & $\ell_1$ SNR$={30}$~dB \\ 
 \includegraphics[scale=0.24]{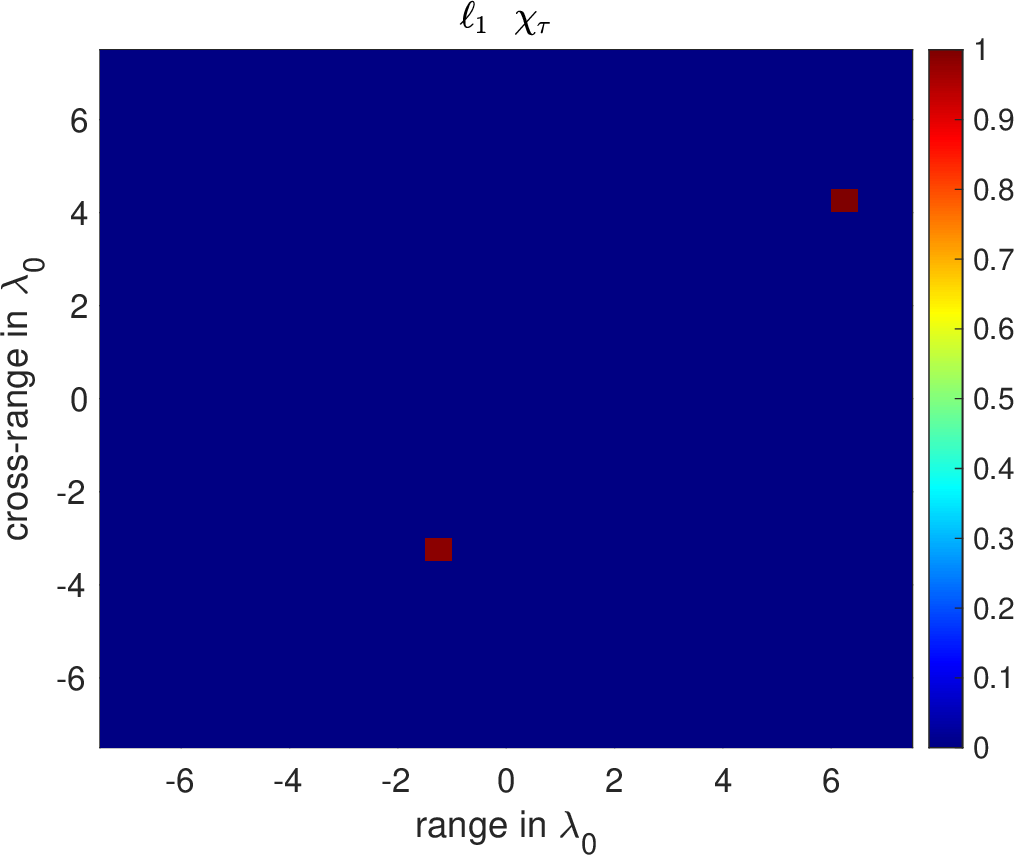} &
 \includegraphics[scale=0.24]{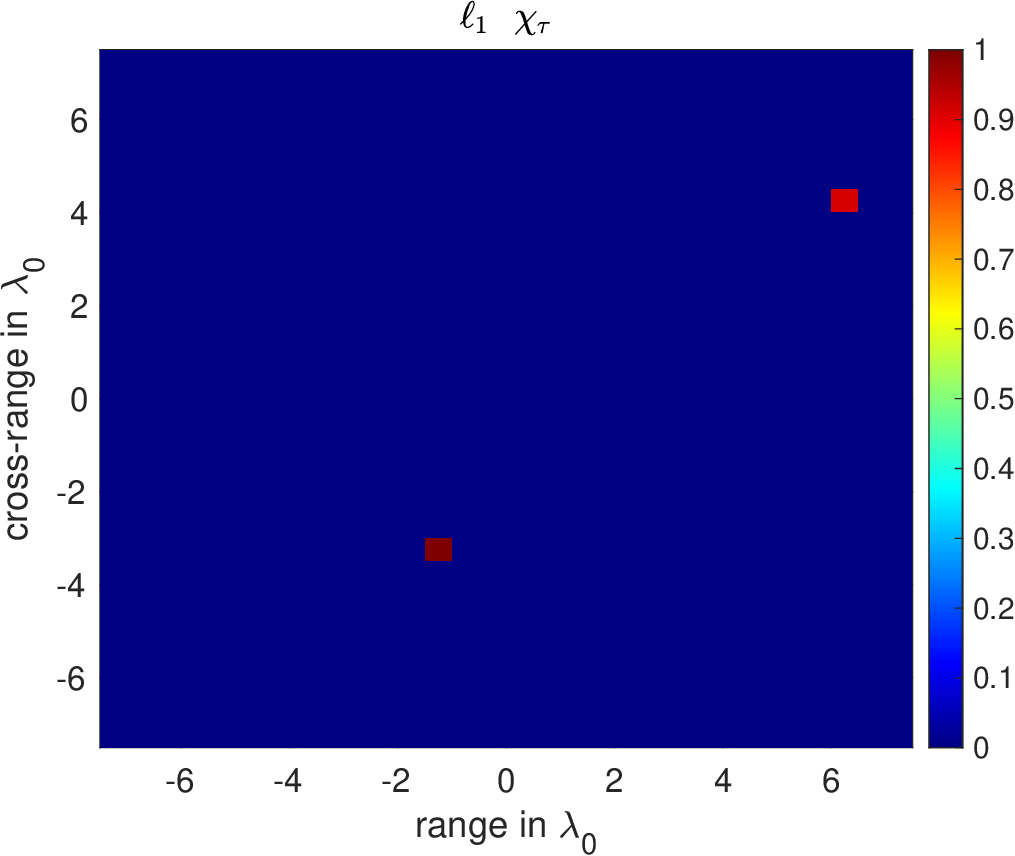}  \\
  \includegraphics[scale=0.24]{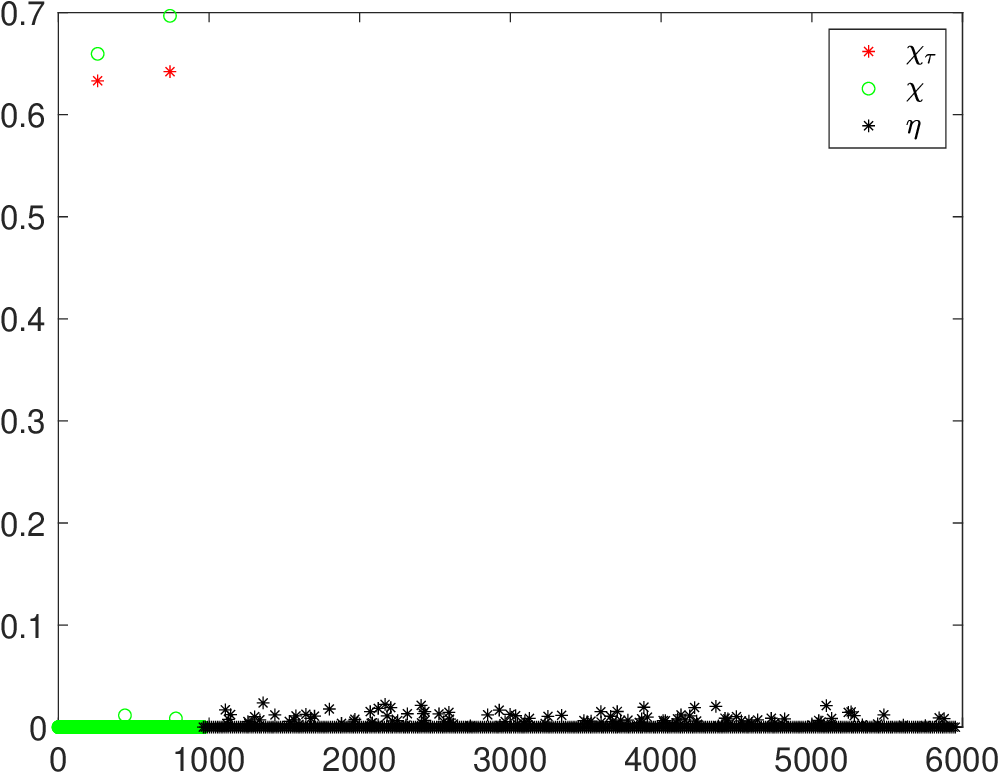} &
 \includegraphics[scale=0.24]{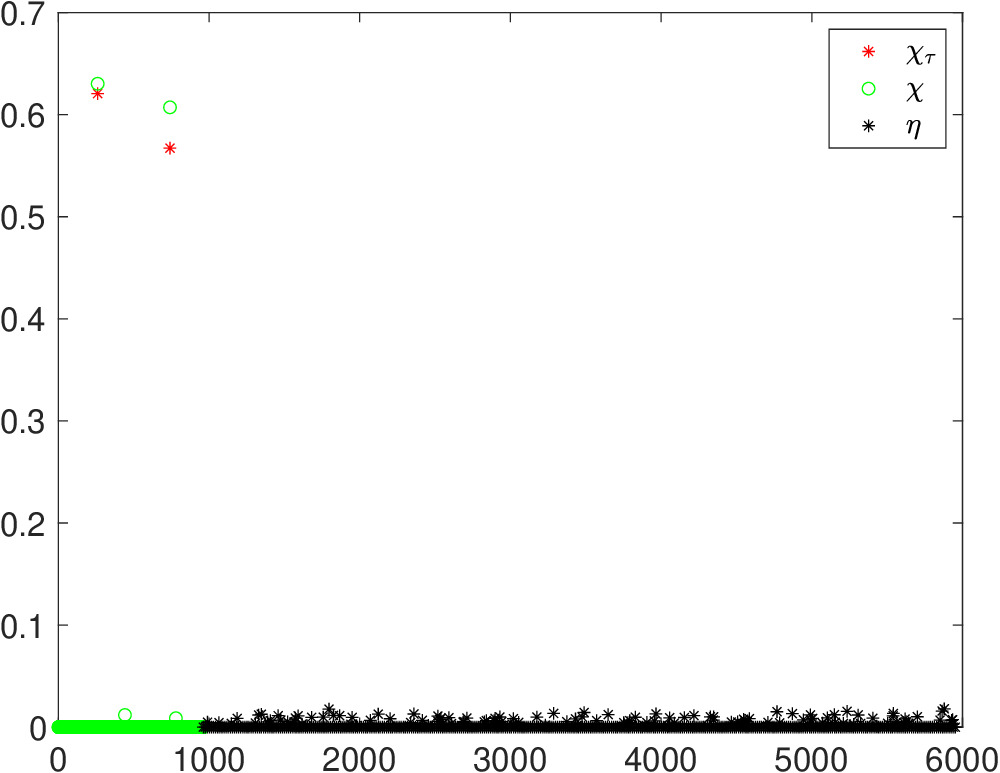} \\
   \end{tabular}
\end{center}
 \caption{First step of the algorithm. Imaging $M=4$ absorbing objects using the total power received on the array from $N=300$ illumination patterns.  The top plot shows the true distribution of absorbers; two are strong (red squares) and two are weak (white crosses).  Bottom panel of four figures: reconstructions with no noise (left column) and with additive noise so the SNR is $30$dB (right column). 
 } 
 \label{fig1} 
 \end{figure}

In the first step of the algorithm, we seek to reconstruct the strong absorbing objects. As explained in section \ref{sec:algo}, we set ${\cal A}={\cal W}_{incoh}$, and solve 
$$
{\cal A} \vect \chi + {\cal C}\vect \eta = \vect d\, ,
$$
for $\vect \chi=[|\rho_1|^2,|\rho_2|^2,\dots,|\rho_K|^2]^T$. The noise collector term ${\cal C}\vect \eta$ absorbs the contributions of $\rho^*_i\rho_j$ for $i \ne j$ to the data which are neglected in this step and treated as noise. Looking at 
$$
\dsp |(\vect b_i)_s|^2 
= \dsp \underbrace{\sum_{k=1}^K  |w_{ik}|^2 |\rho_k|^2}_{\mbox{indep of s}} +
\sum_{k=1}^m  \sum_{\substack{k'=1\\k'\neq k}}^K  F_{sk} F_{sk'}^*w_{ik} w_{ik'}^* \rho_k \rho_{k'}^*\,  
$$
we observe that the first term is independent of the receiver $s$. Therefore in this first step we use the total intensity as data (the sum over all the receivers).

\begin{figure}[htbp]
\begin{center}
 \begin{tabular}{cc}
 $\ell_1$ no noise & $\ell_1$ SNR=$30$dB \\
 \includegraphics[scale=0.24]{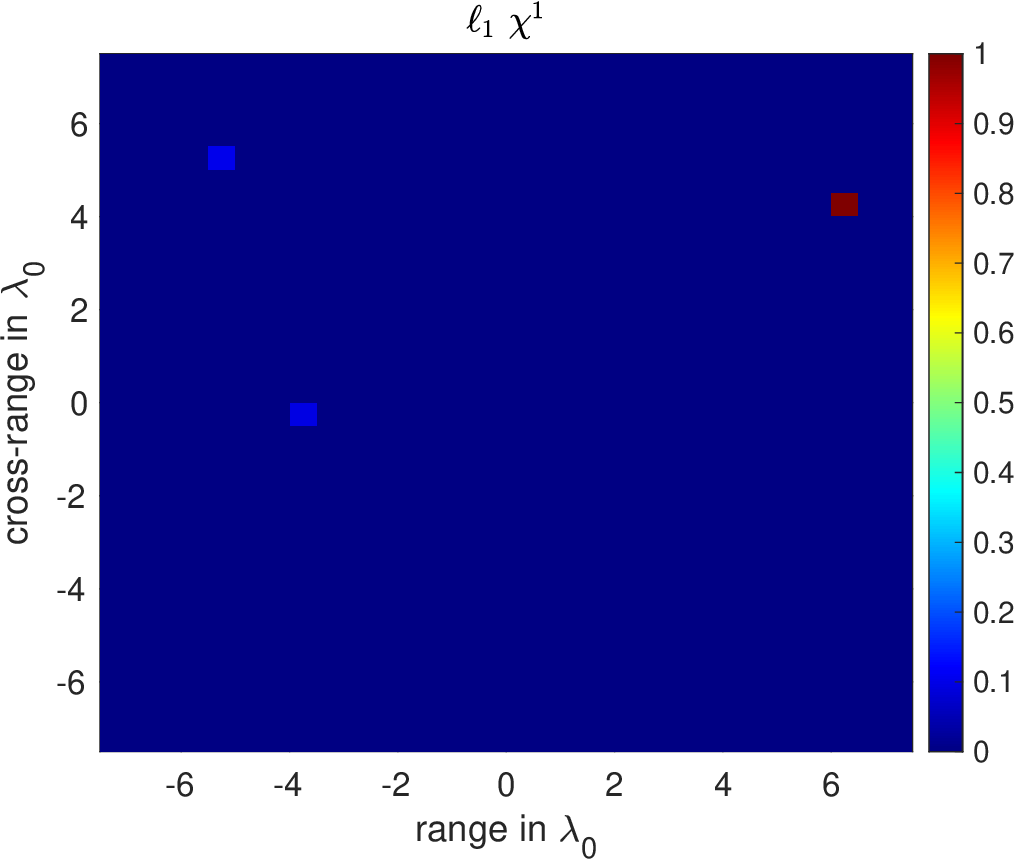}  &  \includegraphics[scale=0.24]{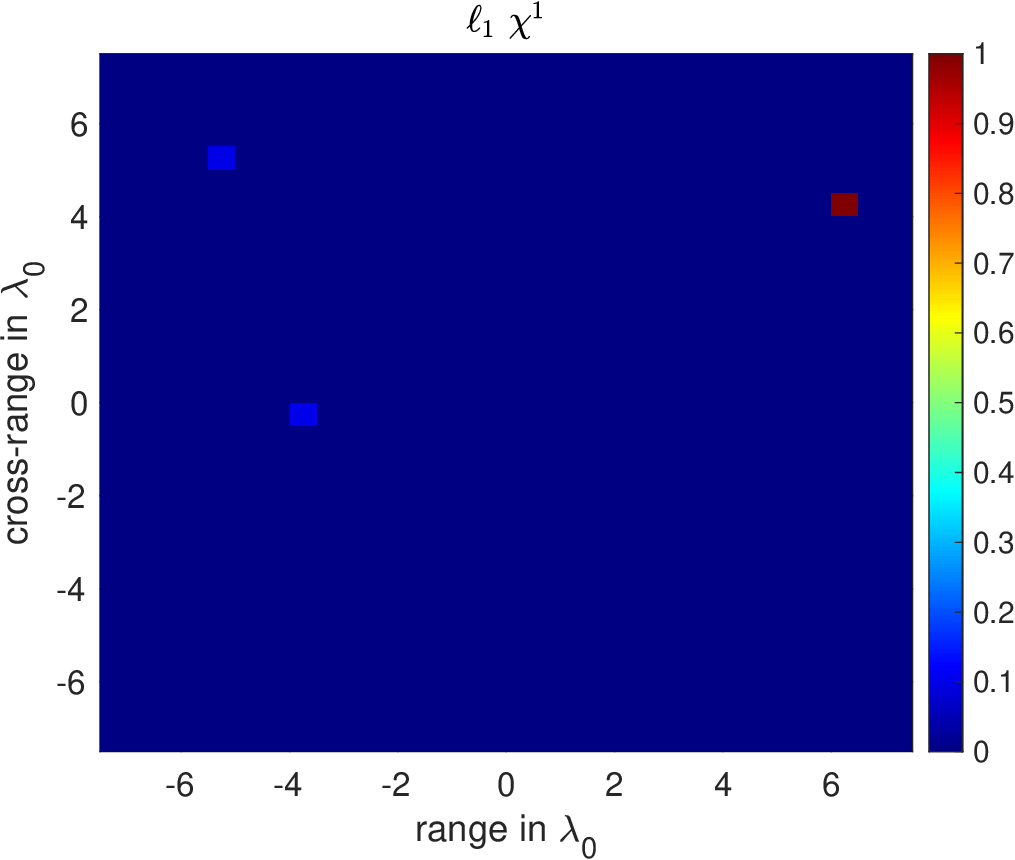}\\
 \includegraphics[scale=0.24]{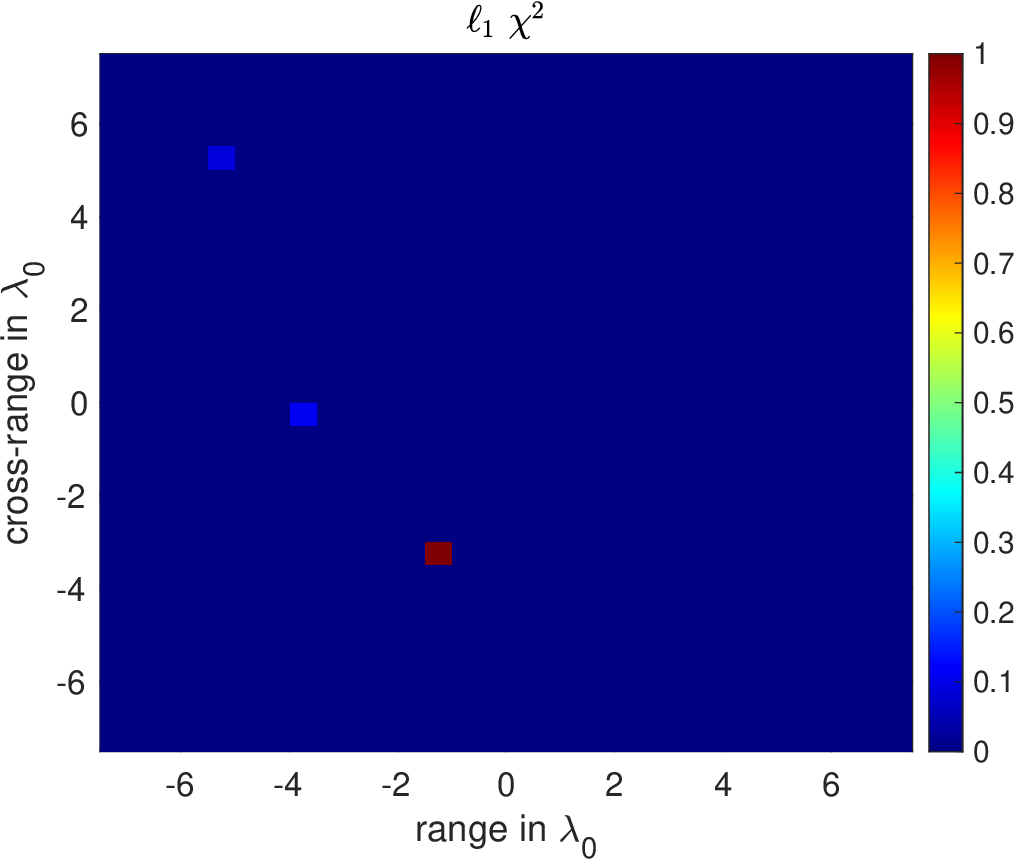} &  \includegraphics[scale=0.24]{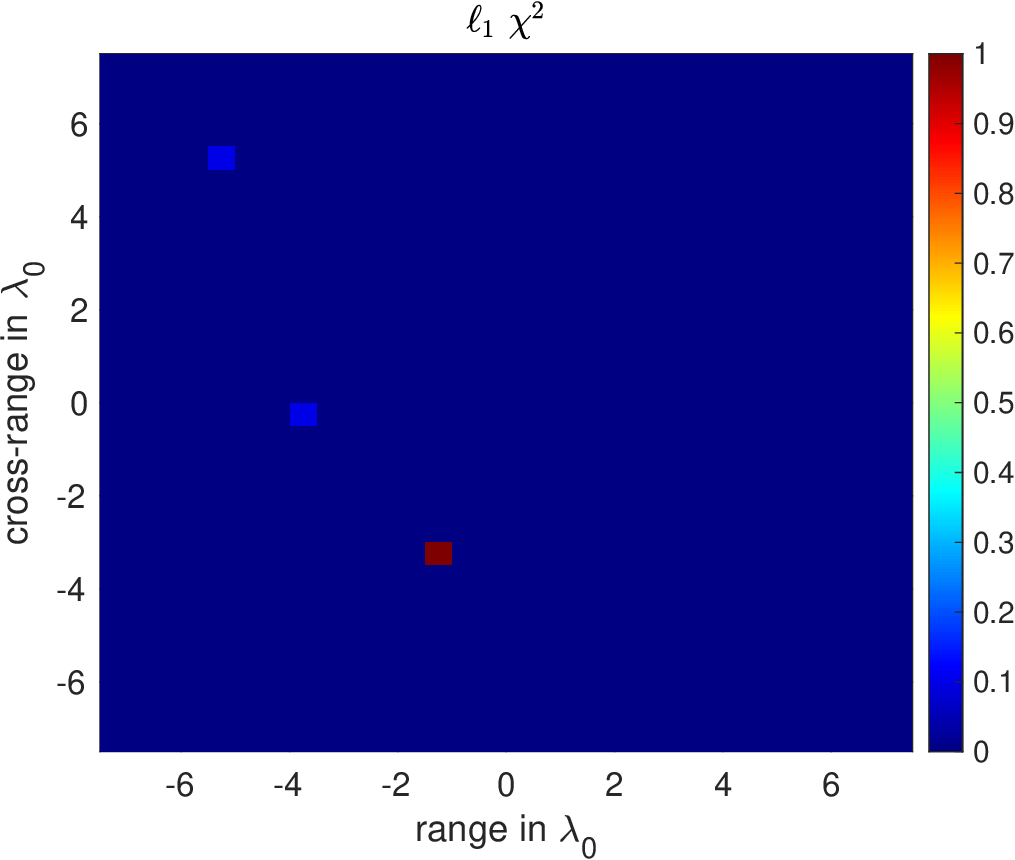} \\
\includegraphics[scale=0.24]{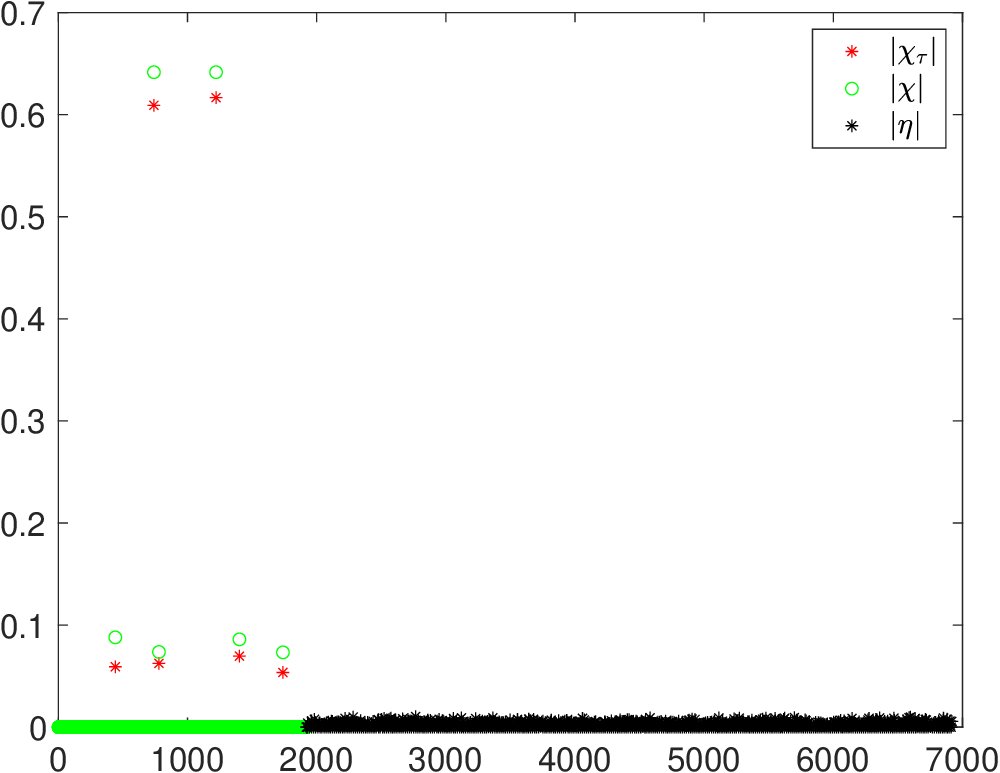}  & \includegraphics[scale=0.24]{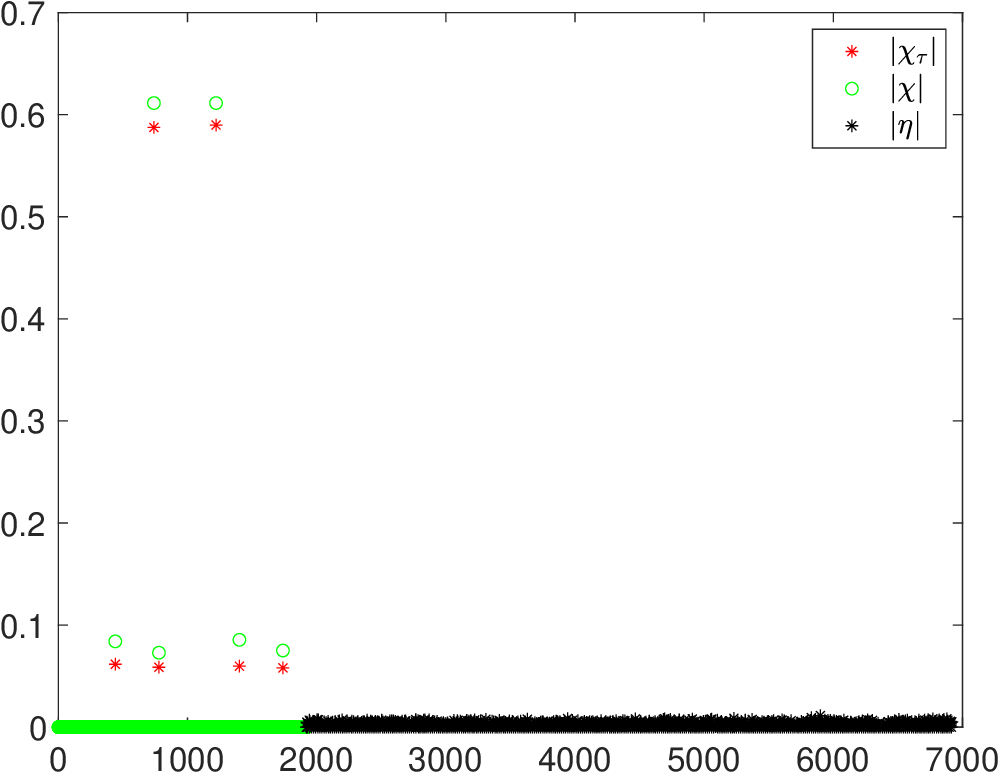} \\
    \end{tabular}\\
\end{center}
 \caption{Second step. Imaging $M=4$ scatterers using intensity measurements over the receiving array.  Left column:  noiseless data. Right column:  SNR $=30$dB. 
 } 
 \label{fig2} 
 \end{figure}

To gain some intuition in how the algorithm works, consider that we want to image $m$ strong absorbers $| \rho_i | = O(1)$, $i=1,\ldots,m$ and $n$ weak absorbers $| \rho_j | = O(\epsilon)$, $j=1, \ldots, n$.  The $n$ weak absorbers have phase contrast that we wish to reconstruct. During the first step of the algorithm we only recover the strong absorbers $| \rho_i |^2$,  $i=1,\ldots,m$, because the contribution from the weak ones $| \rho_j |^2=O(\epsilon^2)$  $j=1, \ldots, n$, is lost in the noise. 

This is illustrated in Figure \ref{fig1}, where we have considered four scatterers, two of them are strong absorbers ($m=2$) shown with red squares and the other two are weak ($n=2$) shown with white crosses.   
The ratio between the weak and the strong absorbing objects in $|\rho_i|^2$ is, in this case, of the order of $1/100$, making the detection of the weak absorbers very difficult.  
The left column of Figure \ref{fig1} are the results for noise-free data, and the right column is the results for data with  SNR$=30$~dB. In both cases, the locations of the strong absorbing objects are recovered exactly. Moreover, their amplitudes $|\rho_i|^2$ are recovered with a quite good accuracy. In the first and second rows of Figure \ref{fig1} we display $|\rho|^2$ as a two-dimensional image, while in the third row we plot $| \rho |^2$ as a vector. In this third row, we plot the exact $| \rho |^2$ vector with green circles, and the recovered one with red stars. The black stars are the non-physical unknown $\vect \eta$ introduced in the algorithm to absorb the contribution to the data due to the cross-terms $\rho_i^* \rho_j$, with $i \neq j$. In both cases, with or without noise in the data, the weakly absorbing objects are not recovered because the neglected contribution to the data of the cross-terms $\rho_i^* \rho_j$, with $i \neq j$ , is larger than the contribution of the weakly absorbing objects (these are the second term which is $O(1)$ and third term which is $O(\epsilon)$ in (\ref{eq:epsilon})).

To find the weak absorbers we apply the second step of the algorithm. To this end, we first remove  from the data the $O(1)$ contributions $\sum_{k=1}^m |w_{ik}|^2 |\rho_k|^2$ from the $m$ strong absorbers  already found in step $1$. 
For our model problem with $m$-strong and $n$-weak absorbers what remains is 
\begin{equation}
\begin{array}{lll}
  &\dsp  \underbrace{{\sum_{k=1}^n |w_{ik}|^2 |\rho_k|^2}}_{O(\epsilon^2)}  +
  \underbrace{\sum_{k=1}^m\sum_{\substack{k'=1\\k'\neq k}}^m  F_{sk} F_{sk'}^*w_{ik} w_{ik'}^* \rho_k \rho_{k'}^*}_{O(1)} \\
   &
   + \underbrace{\sum_{k=1}^m\sum_{\substack{k'=1\\k'\neq k}}^n  F_{sk} F_{sk'}^*w_{ik} w_{ik'}^* \rho_k \rho_{k'}^*}_{O(\epsilon)} +
     \underbrace{\sum_{k=1}^n\sum_{\substack{k'=1\\k'\neq k}}^n  F_{sk} F_{sk'}^*w_{ik} w_{ik'}^* \rho_k \rho_{k'}^*}_{O(\epsilon^2)} 
   \end{array}
   \label{eq:epsilon}
\end{equation}

Then, for every pixel $i=1,\ldots,m$ detected during the first step we seek for its interactions $\rho_i^* \rho_j$ with all the other $K-1$ pixels in the object plane, $j=1, \ldots, K$, $j\neq i$. These are the $O(1)$  and $O(\epsilon)$ contributions to the data (second and third term in (\ref{eq:epsilon})). In this case ${\cal A}=({\cal W}_{coh})_{sub}$, where $({\cal W}_{coh})_{sub}$ contains  the $m\,(K-1)$ columns that correspond to the interactions between the $m$ detected objects in the first step and all the other pixels in the image. Since we are neglecting the {\em $O(\epsilon^2)$ contributions}, the system is not exact. 

\begin{figure}[htbp]
\begin{center}
 \begin{tabular}{cc}
$\ell_1$ no noise & $\ell_1$ SNR=$30$dB\\
 \includegraphics[scale=0.24]{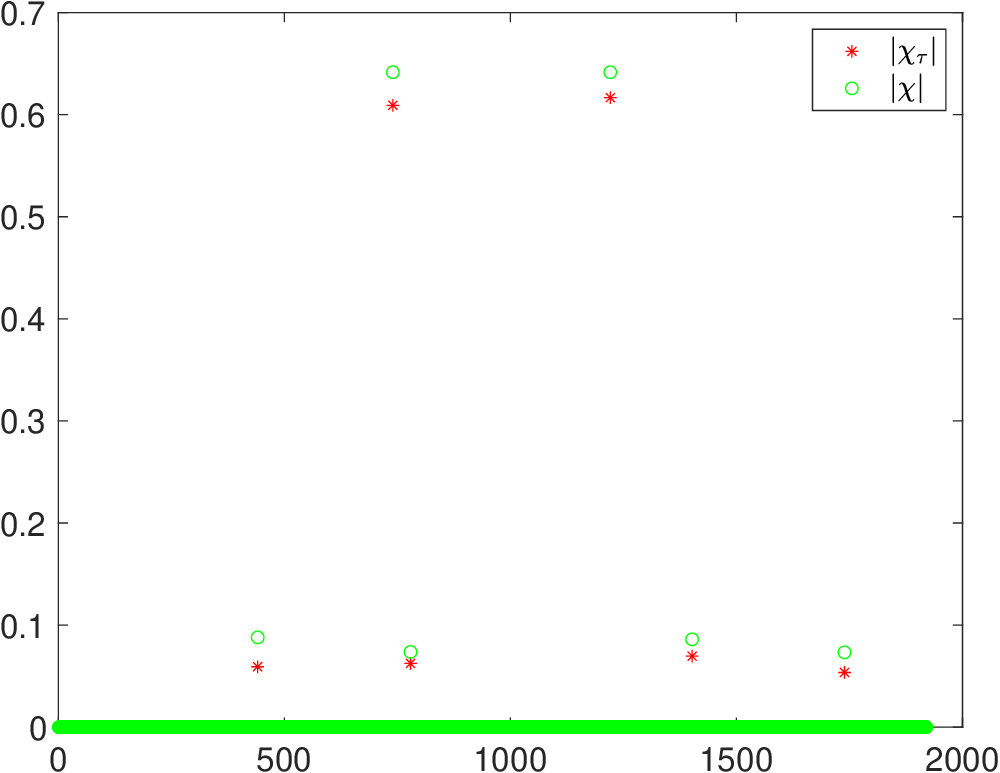} &  \includegraphics[scale=0.24]{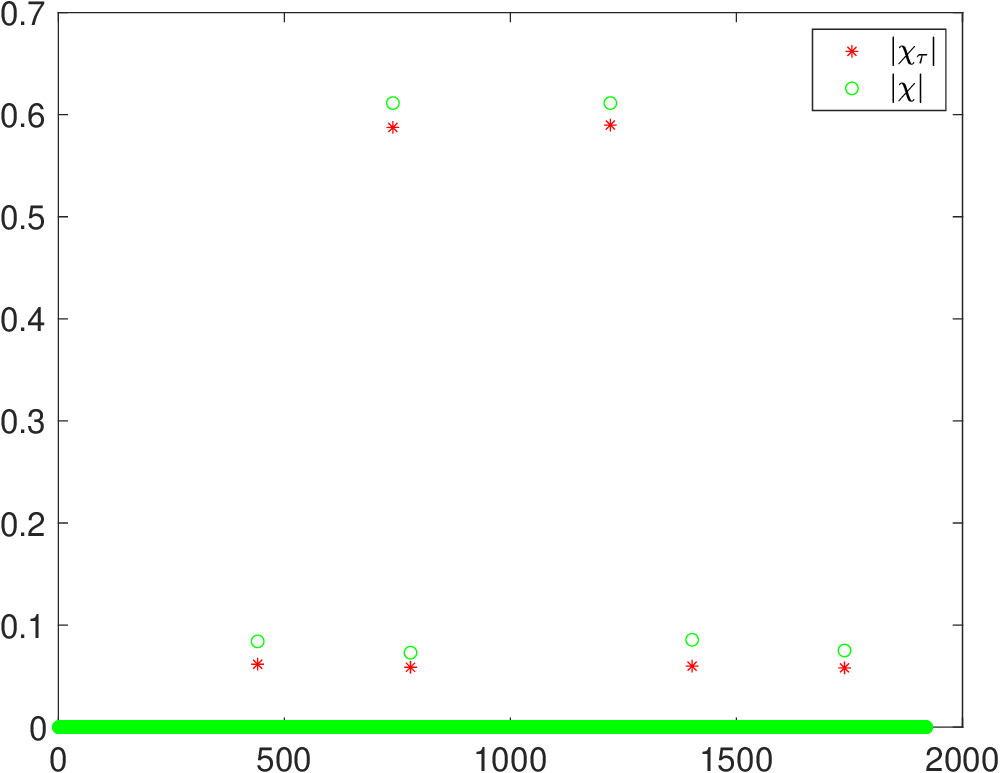}\\
\includegraphics[scale=0.24]{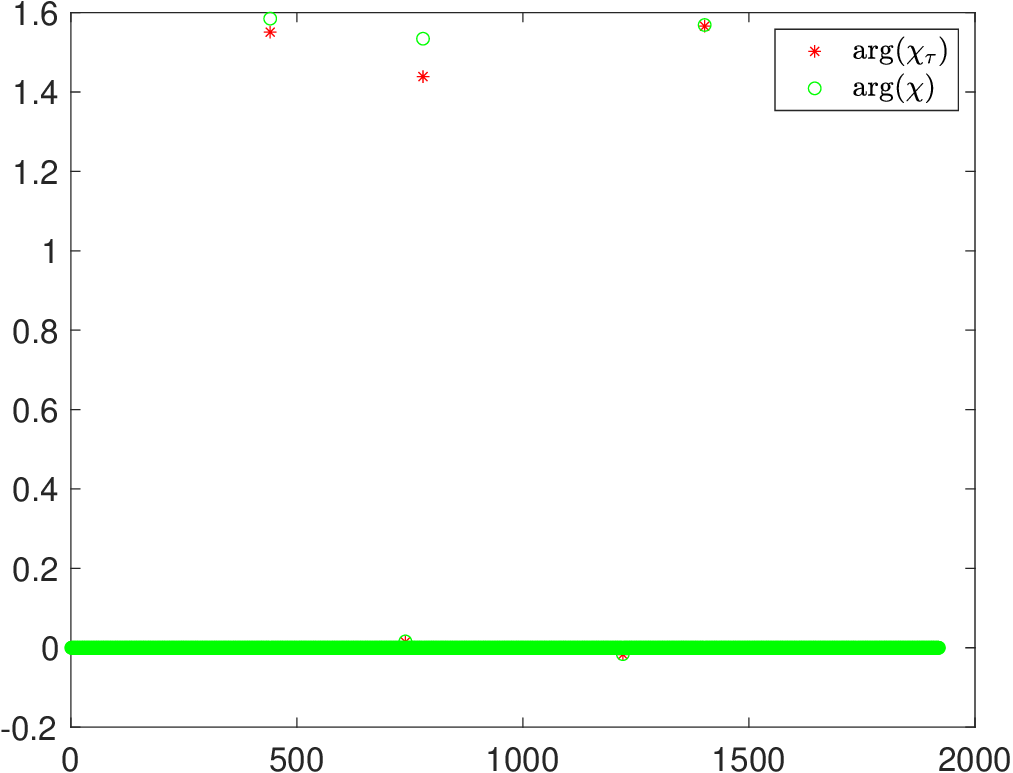}  & \includegraphics[scale=0.24]{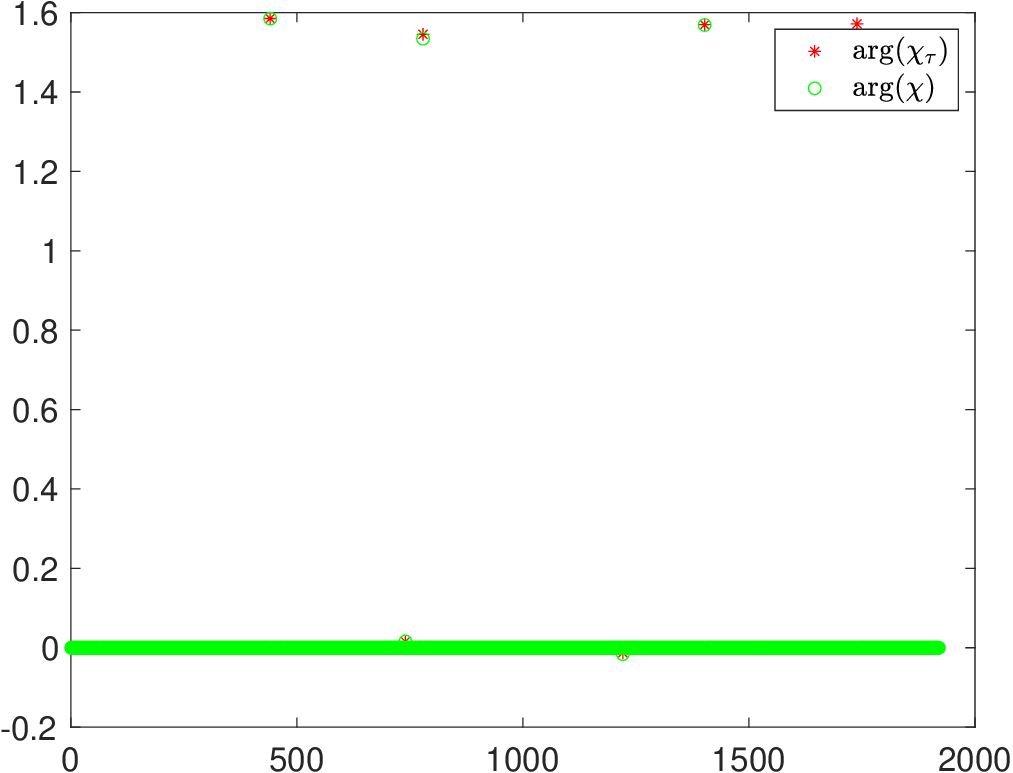}
    \end{tabular}\\
   \end{center}
 \caption{Second step of the algorithm. Left column:  noiseless data. Right column:  SNR $=30$dB. The top row shows the recovered amplitudes and the bottom row the recovered phases. 
 } 
 \label{fig3} 
 \end{figure}

\begin{figure}[htbp]
\begin{center}
true \\[5pt]
 \includegraphics[scale=0.24]{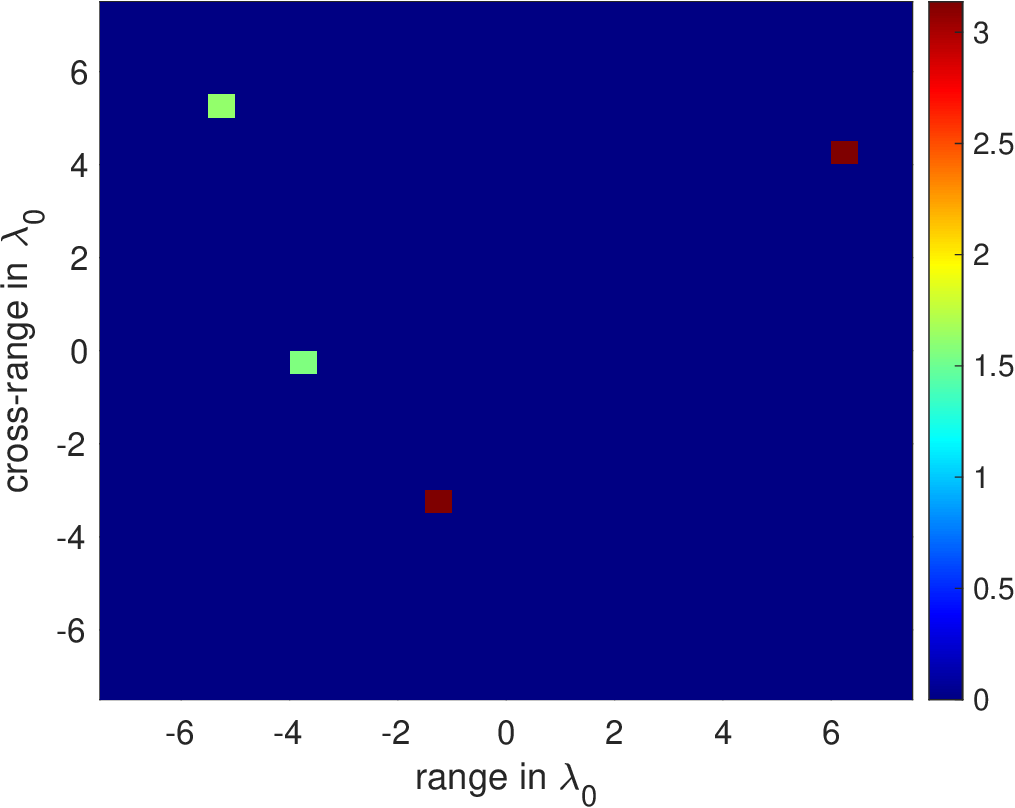}\\
 \begin{tabular}{cc}
  inf SNR & SNR=30dB \\
\includegraphics[scale=0.24]{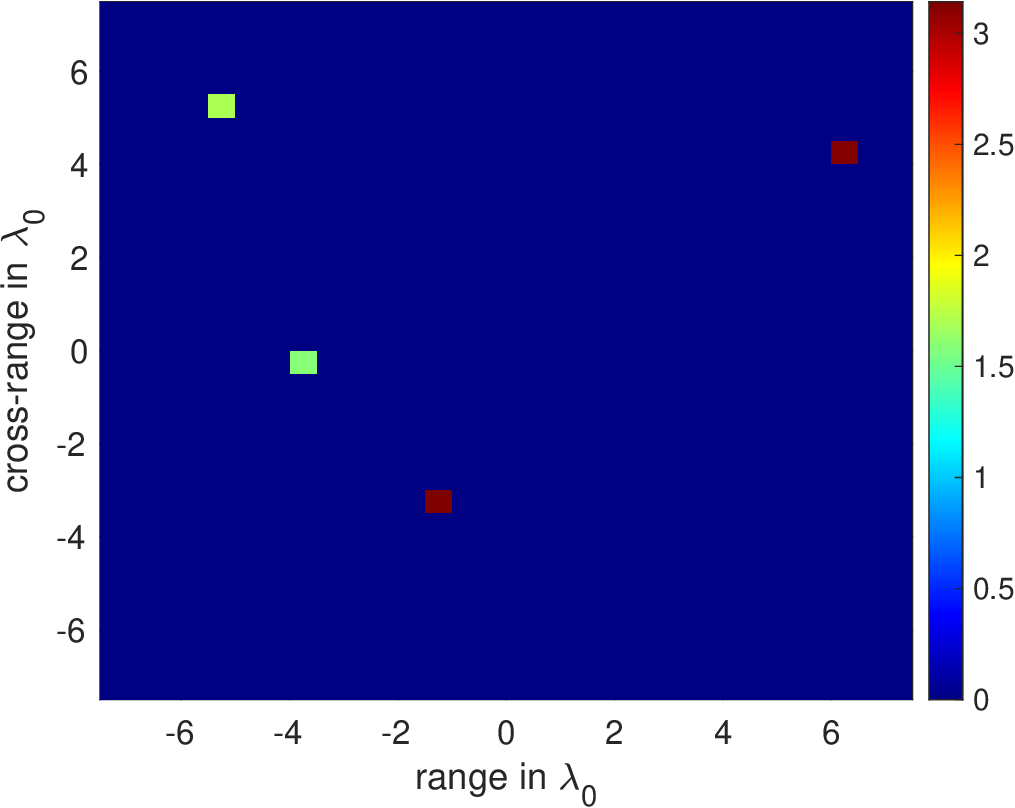} &
 \includegraphics[scale=0.24]{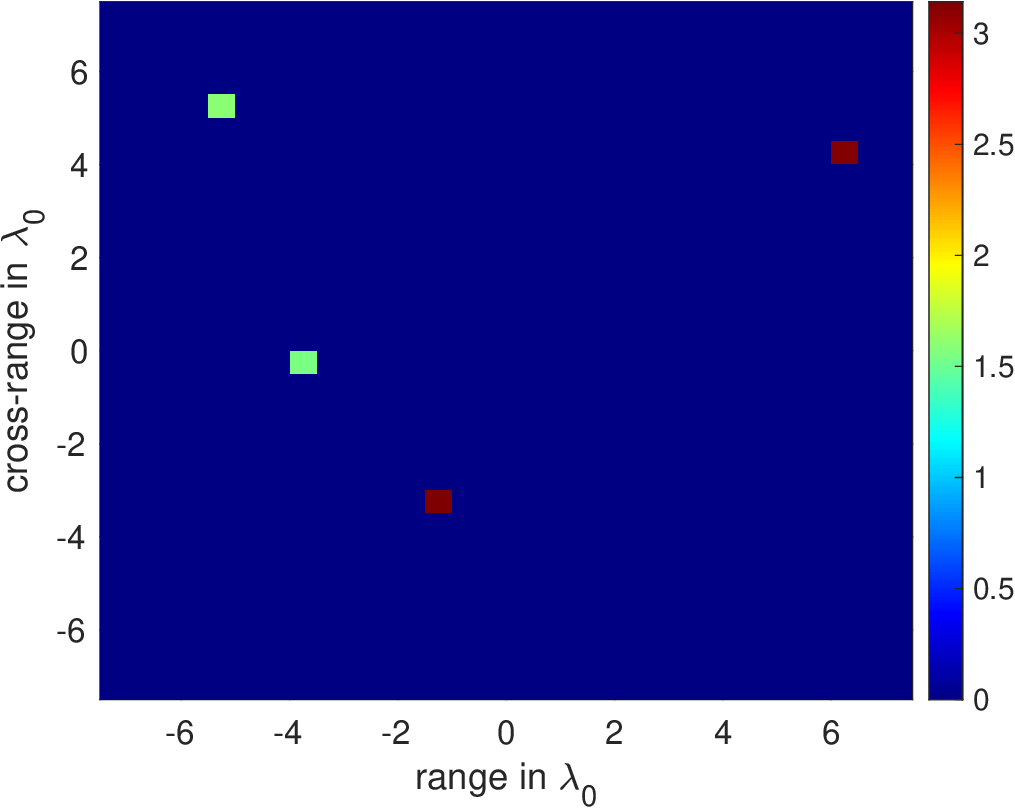} \\
    \end{tabular}
   \end{center}
 \caption{Phase maps corresponding to Figure \ref{fig1}. The top plot is the true phase distribution, and the bottom left and right plots the recovered phase distributions without noise and with noise, respectively.} 
 \label{fig3b} 
 \end{figure}

For the example shown in Figure \ref{fig1}, we found $m=2$ strong absorbers, so we have $2K-2$ unknowns.
The results of this second step are shown in Figure \ref{fig2}. {In top row} we display the unknown recovered by considering the interactions with the first scatterer, while {in the center} row we display the unknown recovered by considering the interactions with the second scatterer. In {the bottom row} we plot the unknown $t_i^* t_j$; the green circles represent the true solution, the red stars the unknown recovered by $\ell_1$ minimization, and the black stars the non-physical part of the unknown corresponding to the {\em Noise Collector.}  This second step finds the weak absorbers with amplitudes and phases recovered with good accuracy; see the results in Figures \ref{fig3} and \ref{fig3b} that summarize these results. In this example we do not  need to apply the third step of the algorithm as both amplitudes and phases of the unknown are recovered correctly after the first two steps.

\begin{figure}[htbp]
\begin{center}  
true $| \rho_k |^2$ \\
  \includegraphics[scale=0.24]{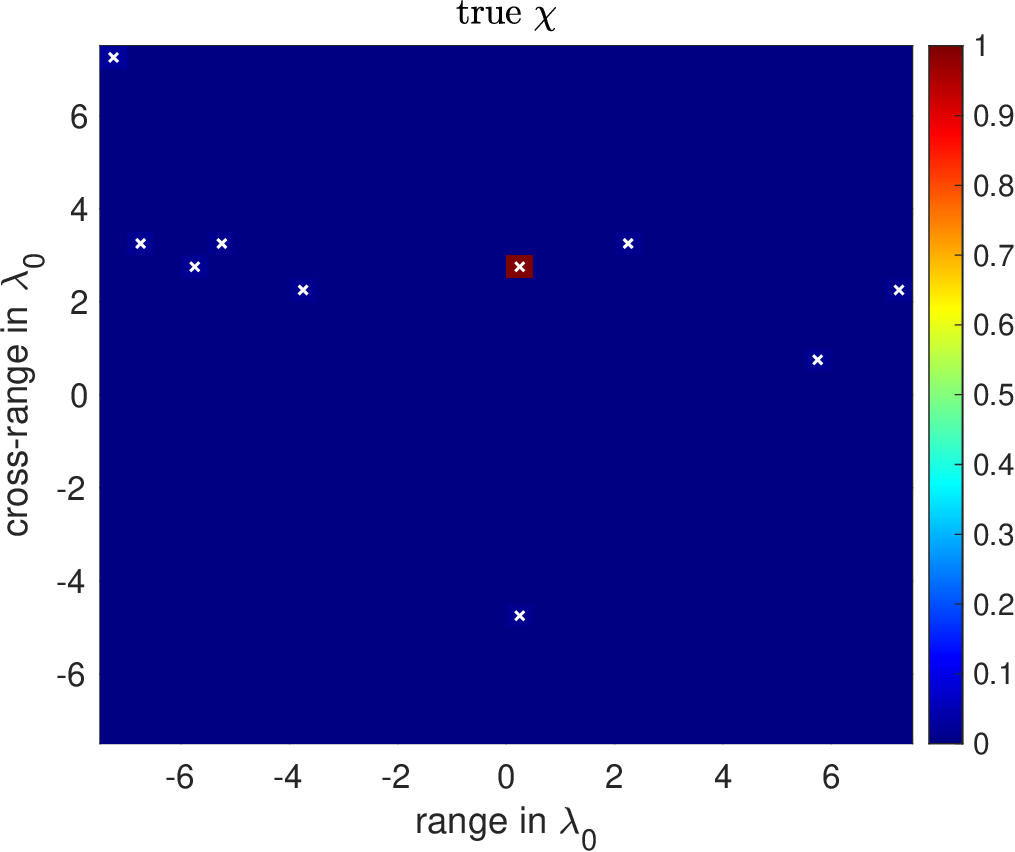} \\
\begin{tabular}{cc}
 $\ell_1$ 1st step & $\ell_1$ 2nd step \\ 
  \includegraphics[scale=0.24]{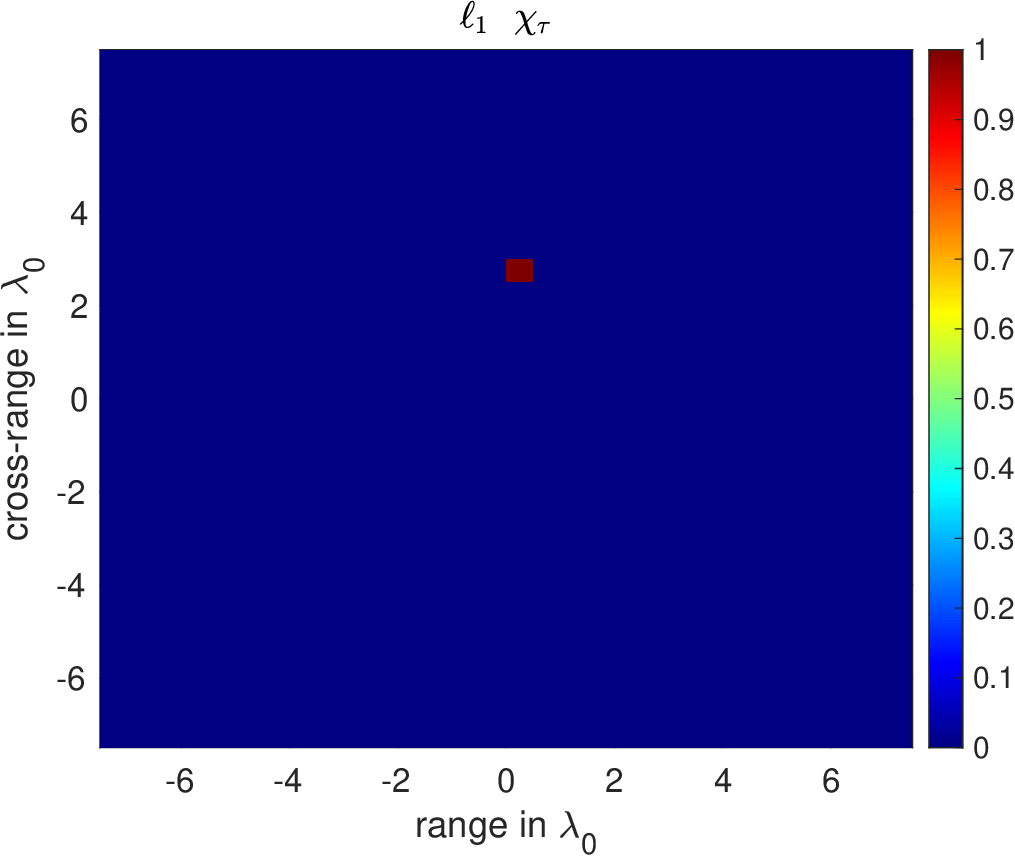} & 
  \includegraphics[scale=0.24]{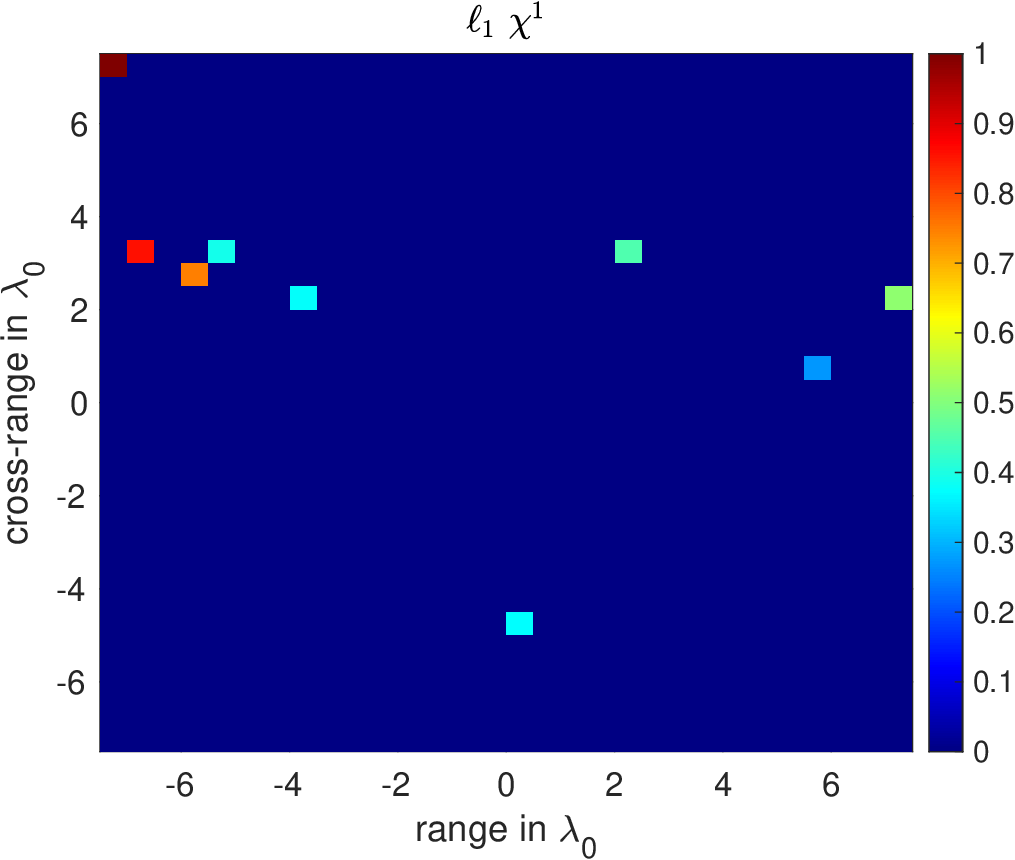} \\
  \end{tabular}
  \\
  \begin{tabular}{cc}
  \includegraphics[scale=0.24]{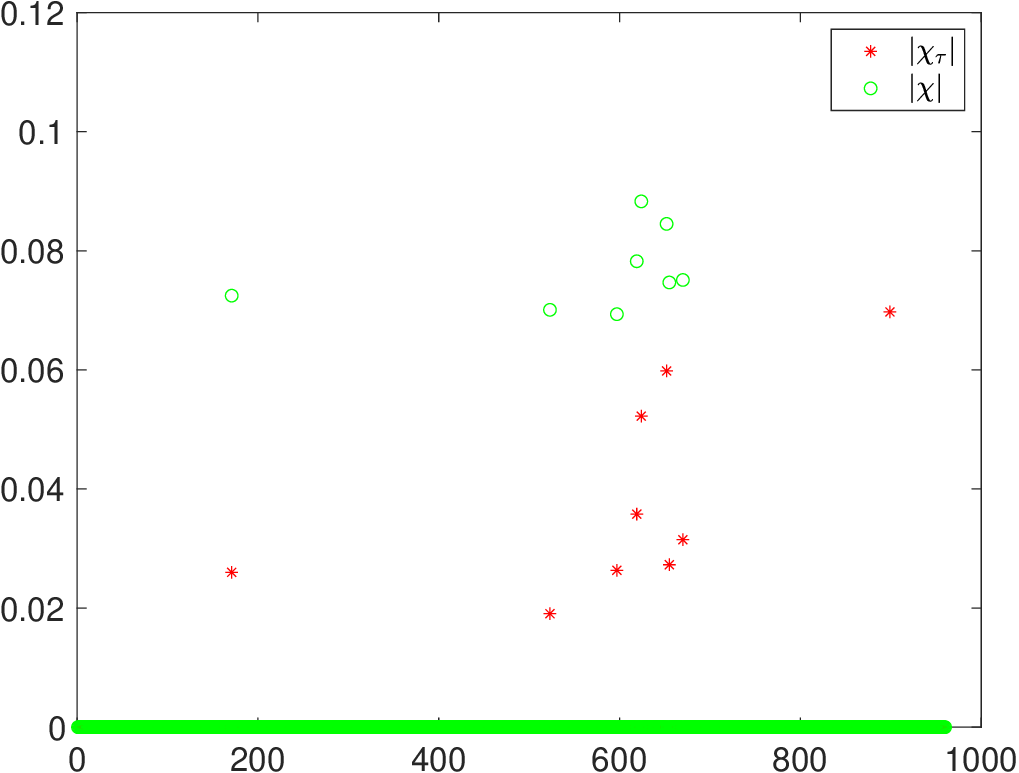} &
  \includegraphics[scale=0.24]{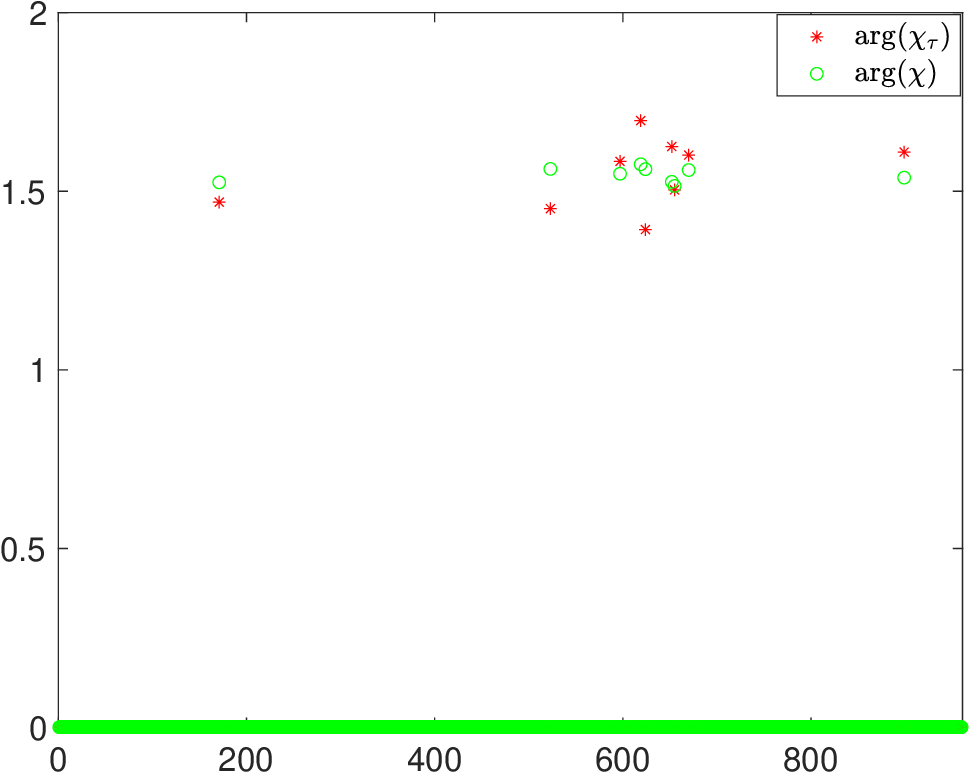} \\
  \end{tabular}
\end{center}
 \caption{First step and second steps for $M=10$ scatterers.  The bottom left and  bottom right plots show the recovered amplitudes and the recovered phases, respectively. SNR$=30$dB.} 
 \label{fig5} 
 \end{figure}

Next, we consider in  Figure \ref{fig5} a more challenging example with ten absorbing objects, with only a strong one ($m=1$ and $n=9$), and $\mbox{SNR}=30$dB.  As in the previous example,  only the strong absorber is found with the first step of the algorithm in the absorption-based image. In the second step, as expected, we also find all the weak ones in the phase contrast image.
The last row of Figure  \ref{fig5} shows the reconstructed amplitudes (left) and phases (right) of the weakly absorbing objects plotted as vectors. The exact values are represented with green circles and the reconstructed values with red stars.

We observe from the results in Figure \ref{fig5} that the phases of the weakly absorbing objects are recovered with much more accuracy than their amplitudes. This is because the error induced by the neglected terms in the second step increases with the number of objects. Indeed, in the second step we  only account for the interactions between the strong and the weak absorbers (the $O(\epsilon)$ term in (\ref{eq:epsilon}) that corresponds to $n=9$ contributions here), but we neglect all the interactions between the weak absorbers (this is the last $O(\epsilon^2)$  term in (\ref{eq:epsilon})  which corresponds to $n^2=81$ contributions here). This example is therefore more challenging because the modelling error increases quadratically with the number of weak absorbers. 

\begin{figure}[htbp]
\begin{center}
 \begin{tabular}{cc}
   \includegraphics[scale=0.24]{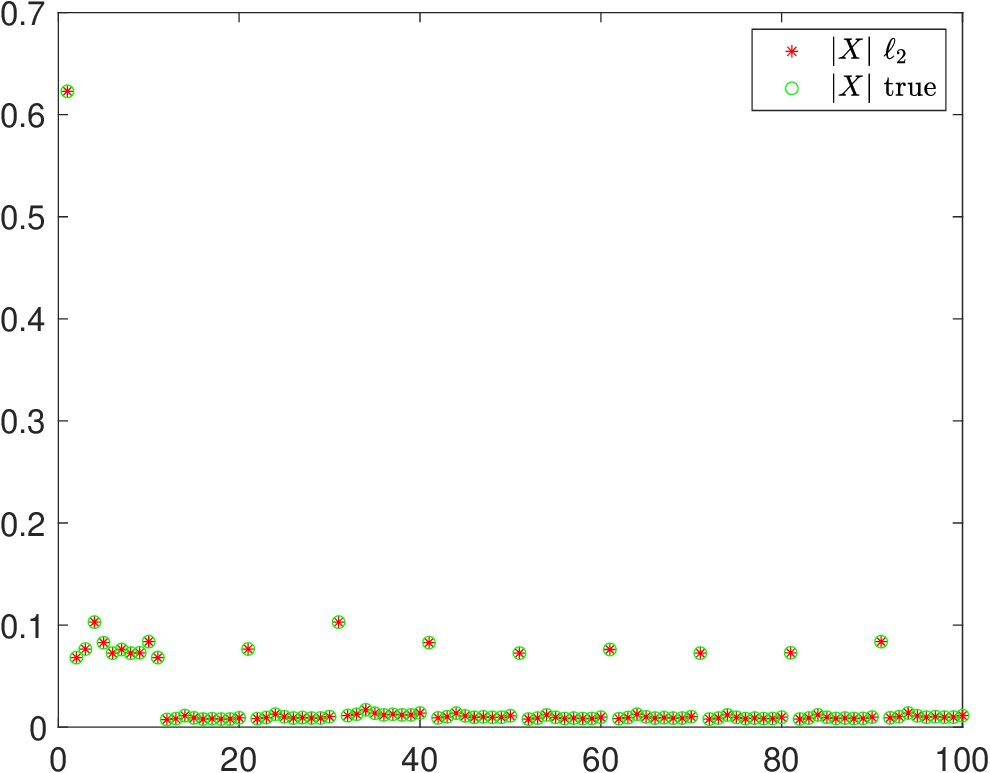} &
  \includegraphics[scale=0.24]{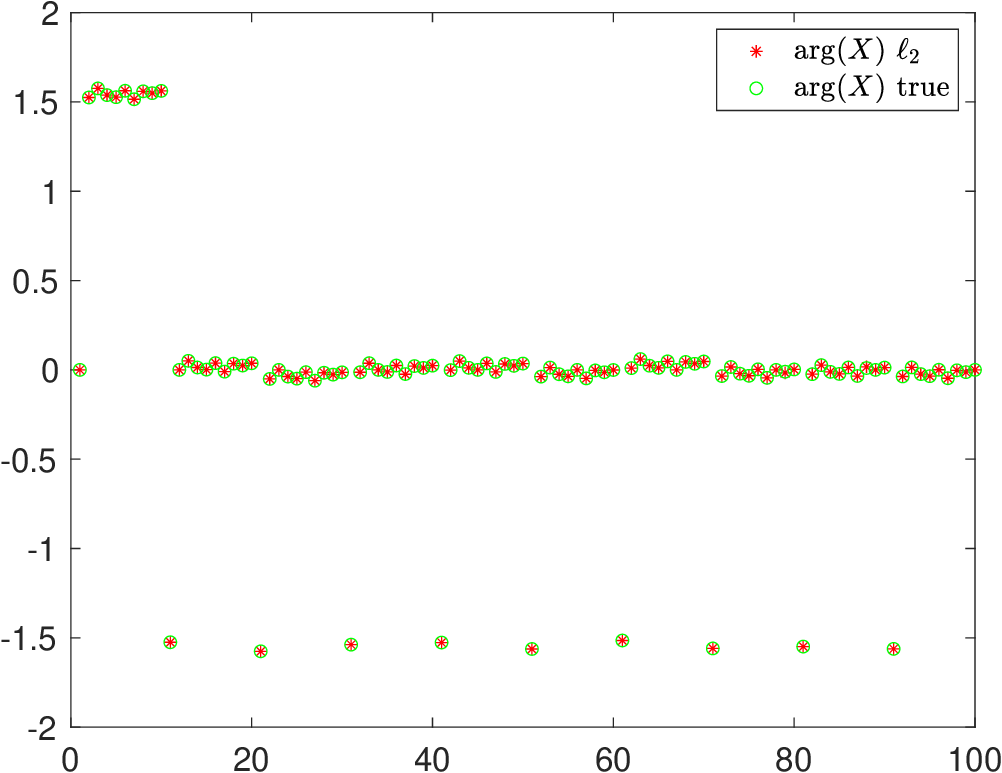} \\
   \end{tabular}
   \end{center}
 \caption{Third step for the full unknown $X=t t^*$ restricted to the recovered support. SNR$=30$dB. 
 } 
 \label{fig6} 
 \end{figure}

\begin{figure}[t]
\begin{center}
 \begin{tabular}{cc}
 true & SNR=30dB \\
 \includegraphics[scale=0.24]{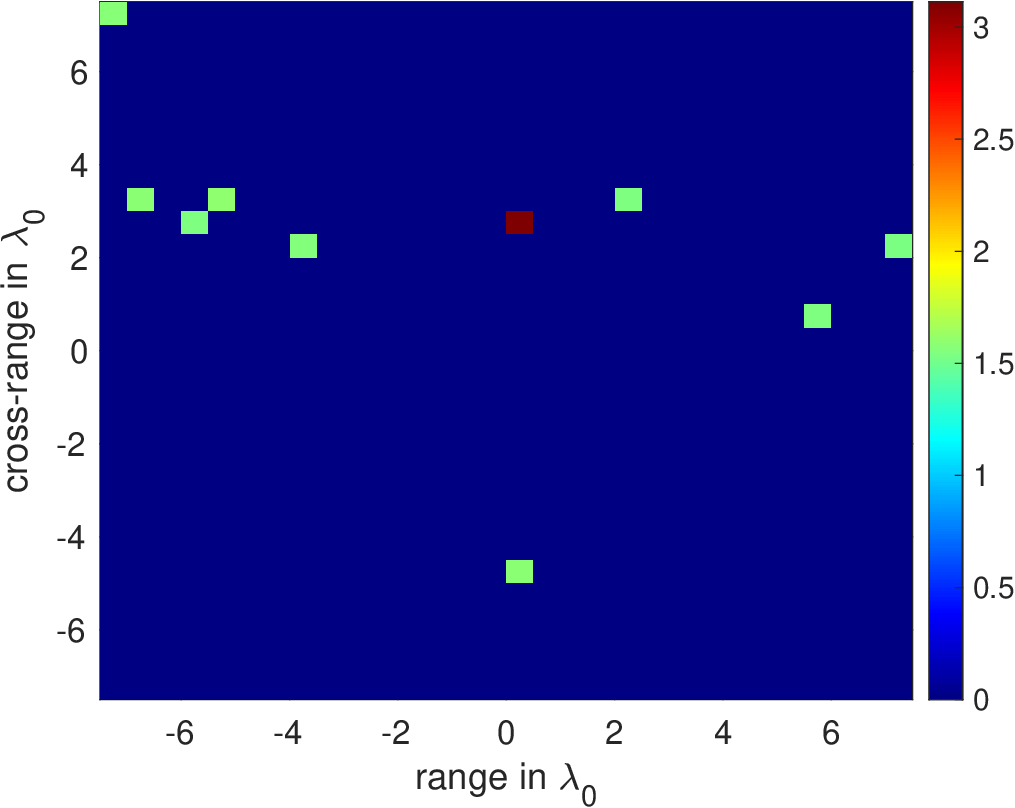} &
\includegraphics[scale=0.24]{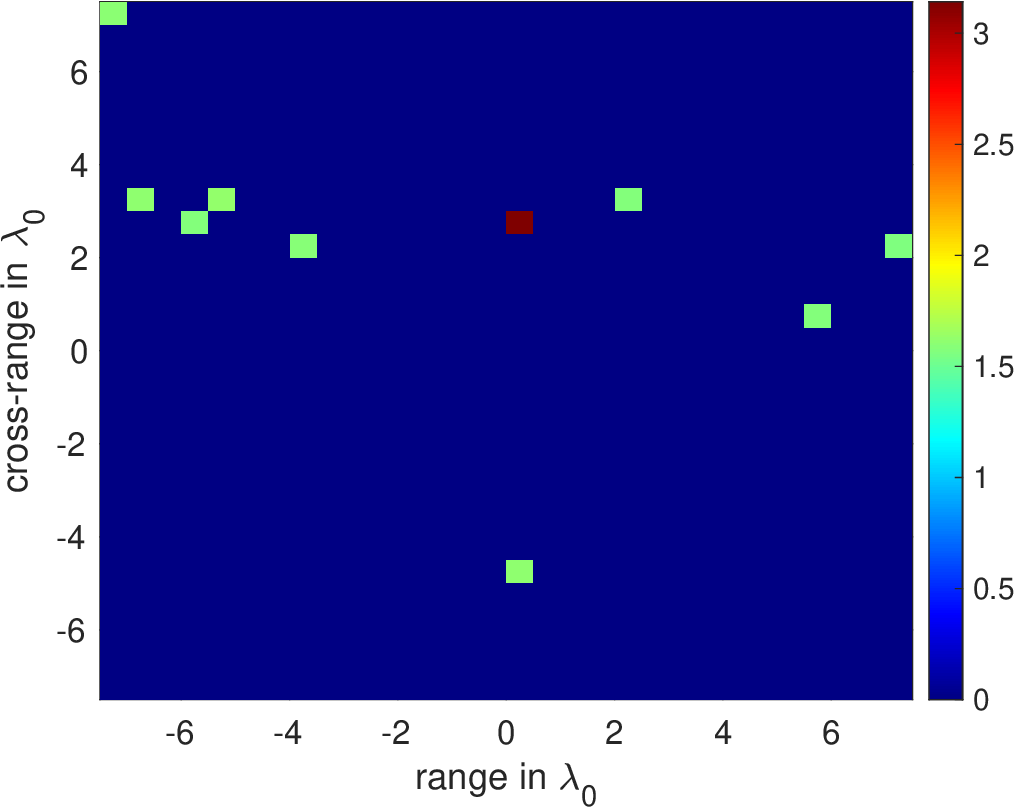} \\
    \end{tabular}
   \end{center}
 \caption{True and recovered phase maps for the $M=10$ absorbers.  SNR$=30$dB. } 
 \label{fig6b} 
 \end{figure}

Better results can be obtained by considering, in a third step, the full problem  \eqref{eq:hugelinearsystem_absorption} restricted to the recovered support. This third step allows us to recover the unknown $X_{supp}=\rho_{supp} \rho_{supp}^*$ accurately using an $\ell_2$ minimization method.  This is because the locations of all the absorbers both weak and strong are recovered exactly after the first two steps of the algorithm. In the third step the contributions to data from all terms in $X$  are taken into account. Figure \ref{fig6} shows that this provides a great accuracy in the recovered values of both the amplitudes and the phases. Figure \ref{fig6b}, that shows the true and recovered phase distributions, illustrates the potential of the proposed imaging method for imaging both strong and weak absorbers with intensity-only measurements. 

Finally, we present results for the case in which the illuminations are partially coherent. This is an interesting case because in some applications fully coherent illuminations are very hard to obtain. We use the following model to generate the data 
\begin{eqnarray}
\label{response:intensity_pcoh}
|(\vect b_i)_s|^2 &=& \sum_{k=1}^K  |w_{ik}|^2 |\rho_k|^2 + \dsp \alpha_{coh} 
\sum_{k=1}^K \sum_{\substack{k'=1\\k'\neq k}}^K  F_{sk} F_{sk'}^*w_{ik} w_{ik'}^* \rho_k \rho_{k'}^*\, ,
\end{eqnarray}
with $0 \le \alpha_{coh} \le 1$. If $\alpha_{coh} = 1$, the sources are fully coherent, and if $\alpha_{coh} = 0$ they are fully incoherent. The parameter $\alpha_{coh}$ that models our uncertainty about the coherence of the illumination is not known when we seek to reconstruct the unknown absorbers.

\begin{figure}[htbp]
\begin{center}  
  \begin{tabular}{cc}
  \includegraphics[scale=0.24]{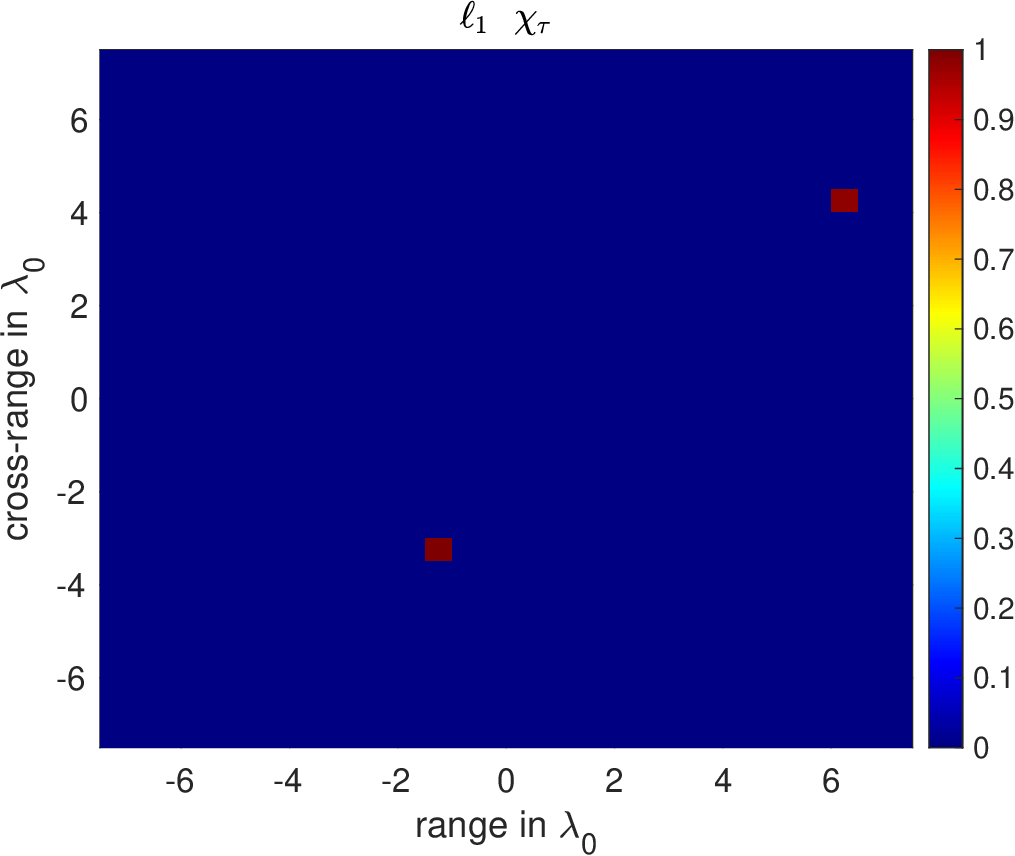} &\includegraphics[scale=0.24]{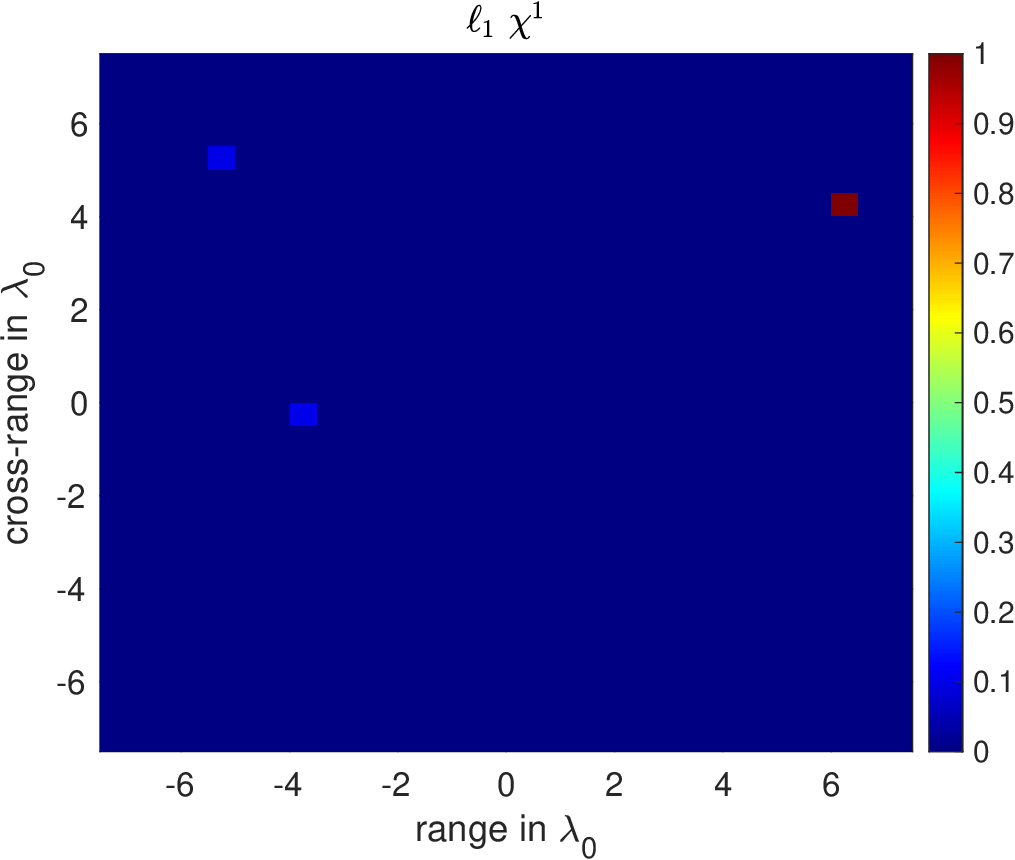}\\
\includegraphics[scale=0.24]{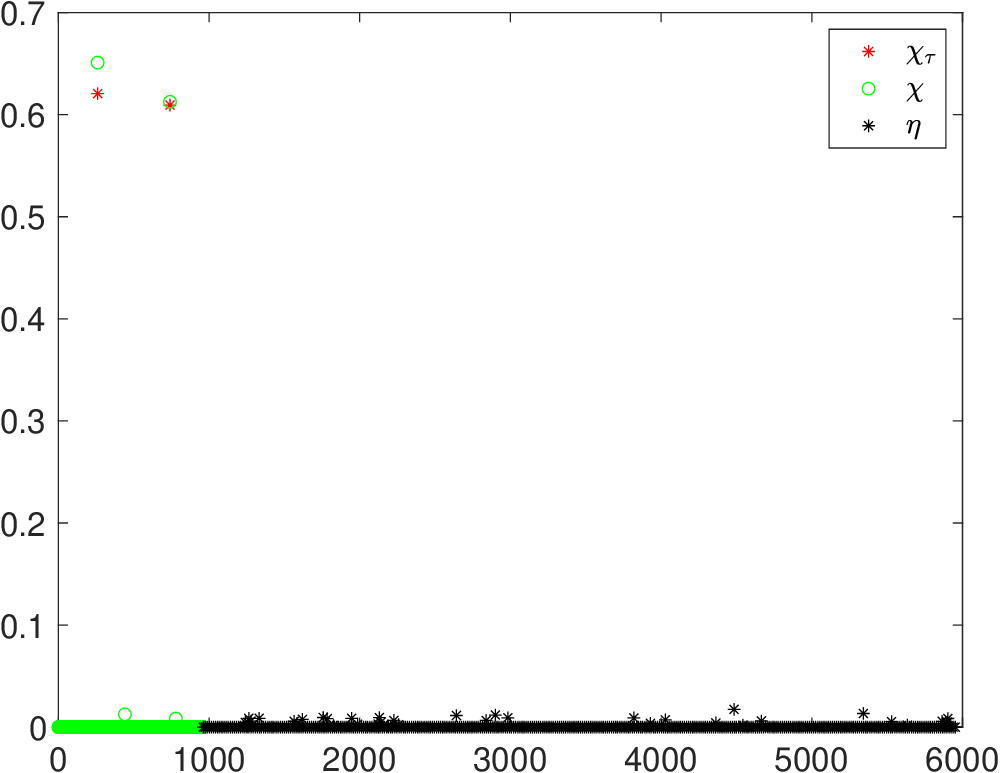}  &  \includegraphics[scale=0.24]{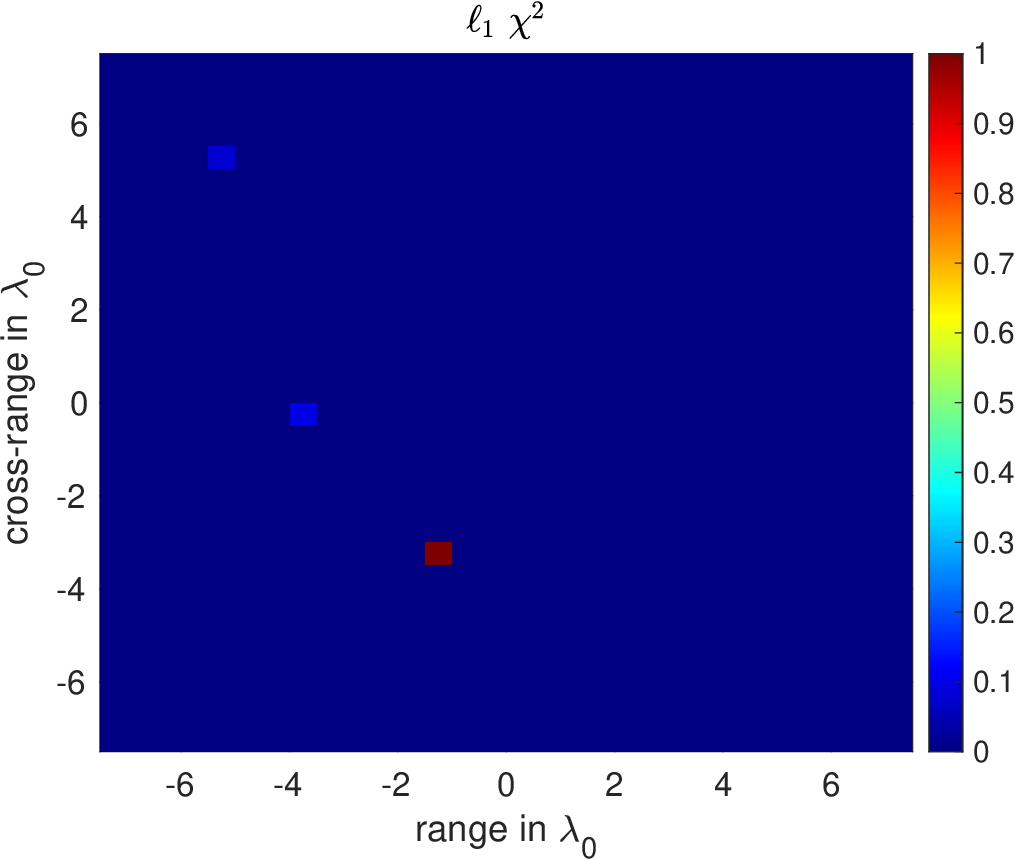} \\
 &\includegraphics[scale=0.24]{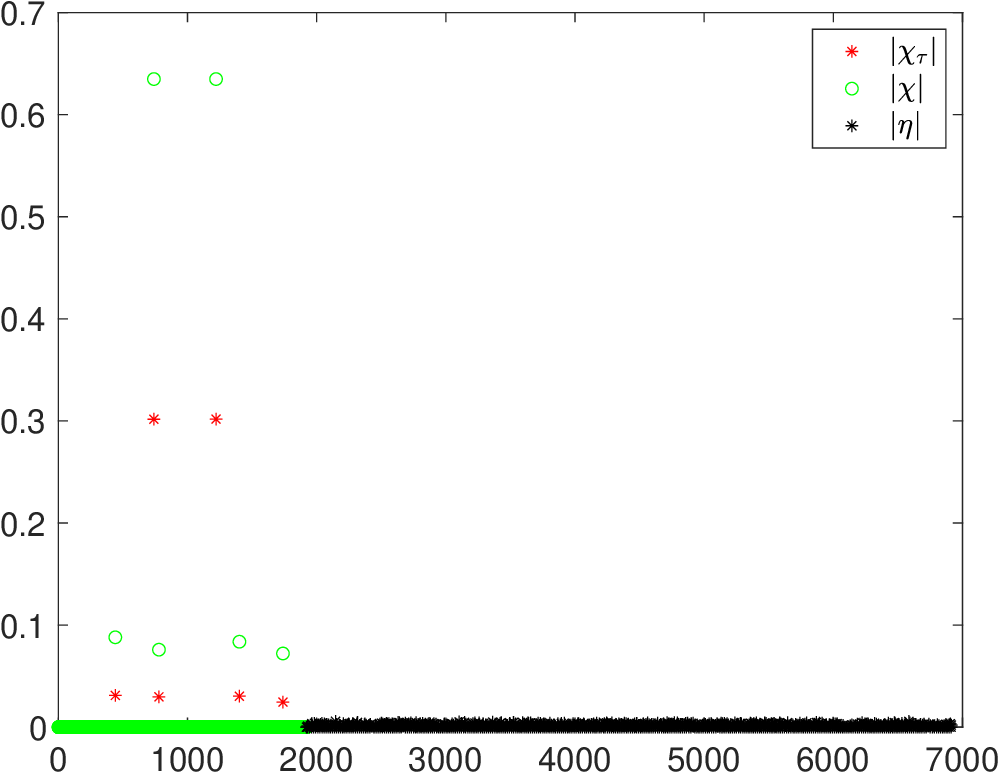} \\
  \end{tabular}
\end{center}
 \caption{Imaging $M=4$  with partially coherent illumination; $\alpha_{coh}=0.5$. Left and right columns: first and second steps of the algorithm, respectively. There is no noise in the data.} 
 \label{fig:incoh} 
 \end{figure}

Figure \ref{fig:incoh} shows the results when the illumination used for imaging is partially coherent; $\alpha_{coh} = 0.5$ in this numerical experiment. The {left and right columns} show the outputs of the first and second steps of the algorithm, respectively. Because the illumination is partially coherent and, thus, the modeling error in the first step is smaller, we observe that the first step recovers the amplitudes of the strong absorbers with great accuracy. The second step is still able to recover the locations of the weak absorbers exactly. However, as expected, we observe that there is a SNR issue, and that if $\alpha_{coh}$ decreases below a certain threshold, we would not be able to image them. This threshold depends on the transparency of these objects, their number, and the noise in the data. For the numerical experiment shown here, the phases of all the absorbers are recovered with the same precision as in the previous experiment (results not shown).

\section{Conclusions}
\label{sec:conclusions}

In this paper we have presented a two steps algorithm for phase retrieval for sparse objects. This algorithm is very efficient because its cost is linear in the number of pixels of the image and, thus, it can be employed for high resolution imaging. It guarantees exact recovery if the image is sparse with respect to a given basis, and it can be used, without any modification, when the illumination is partially coherent. In these cases, only the SNR of the created images is affected. 
Although for ease of presentation this algorithm is introduced for Fourier measurements, it also works for general quadratic measurements without any modification. 
 With this algorithm we are able to image transparent or semi-transparent structures that are not visible in the common used absorption-based images.

\section*{Acknowledgments}
The work of M. Moscoso was partially supported by the grant PID2020-115088RB-I00. The work of A.Novikov was partially supported by NSF DMS-1813943 and AFOSR FA9550-20-1-0026. 
The work of C. Tsogka was partially supported by AFOSR FA9550-21-1-0196.


 \end{document}